\documentclass[%
 aip, 
 reprint,%
]{revtex4-1}

\usepackage[utf8]{inputenc}
\usepackage[T1]{fontenc}
\usepackage{amsmath, amssymb}
\usepackage{bm}% bold math
\usepackage{bbm}
\usepackage{graphicx}% Include figure files
\usepackage[caption=false]{subfig}\usepackage{float}
\usepackage{soul}
\usepackage{dcolumn}% Align table columns on decimal point
\usepackage{mathtools}
\usepackage[bookmarks, colorlinks=true, linkcolor=blue, citecolor=blue, urlcolor=blue, linktoc=page]{hyperref}
\hypersetup{breaklinks=true} 

\usepackage[absolute]{textpos}
\usepackage{xcolor}

\begin{document}
\begin{textblock*}{1.2\textwidth}(2mm, 5mm) % {block width} (coords) 
\centering
\noindent
{\color{green!50!black}  Published in the \href{https://doi.org/10.1103/PhysRevE.108.015306}{\ul{{\it Phys.~Rev.~E}, {\bf 108}, 015306 (2023)}}.
\\[0.5ex]
A copy of the published version can also be obtained by mailing to the authors (emails are at the bottom of this page.)}
\end{textblock*}

%\preprint{AIP/123-QED}
\title[Fundamental solutions of the two-dimensional CCR model]{Fundamental solutions of an extended hydrodynamic model in two dimensions: Derivation, theory, and applications}
% Force line breaks with \\
\author{Himanshi}
\email[Author to whom correspondence should be addressed: ]{phd2001141003@iiti.ac.in}
\affiliation{Department of Mathematics, Indian Institute of Technology Indore, Indore 453552, Madhya Pradesh, India}%Lines break automatically or can be forced with \\
\author{Anirudh Singh Rana}%
\email{anirudh.rana@pilani.bits-pilani.ac.in}
\homepage[Homepage: ]{https://www.bits-pilani.ac.in/pilani/anirudhrana/profile}
\affiliation{ 
Department of Mathematics, Birla Institute of Technology and Science Pilani, Rajasthan 333031, India%\\This line break forced with \textbackslash
%\textbackslash
}%
\author{Vinay Kumar Gupta}
\email{vgk@iiti.ac.in}
\homepage[Homepage: ]{http://people.iiti.ac.in/~vkg/}
\affiliation{Department of Mathematics, Indian Institute of Technology Indore, Indore 453552, Madhya Pradesh, India}
%\author{Vinay Kumar Gupta}
%% \homepage{http://www.Second.institution.edu/~Charlie.Author.}
%\affiliation{%
%Department of Mathematics, Indian Institute of Technology Indore, Indore 453552, India%\\This line break forced% with \\
%}%

\begin{abstract}
The inability of the Navier--Stokes--Fourier equations to capture rarefaction effects motivates us to adopt the extended hydrodynamic equations. 
In the present work, a hydrodynamic model, which consists of the conservation laws closed with the recently propounded coupled constitutive relations (CCR), is utilized.
This model is referred to as the CCR model and is adequate for describing moderately rarefied gas flows. 
A numerical framework based on the method of fundamental solutions is developed to solve the CCR model for rarefied gas flow problems in quasi two dimensions.
%To obtain the solutions for the flow fields, 
To this end, the fundamental solutions of the linearized CCR model are derived in two dimensions. The significance of deriving the two-dimensional fundamental solutions is that they cannot be deduced from their three-dimensional counterparts that do exist in literature.
As applications, the developed numerical framework based on the derived fundamental solutions is used to simulate (i) a rarefied gas flow between two coaxial cylinders with evaporating walls 
and (ii) a temperature-driven rarefied gas flow between two non-coaxial cylinders. 
The results for both problems have been validated against those obtained with the other classical approaches.
Through this, it is shown that the method of fundamental solutions is an efficient tool for addressing quasi two-dimensional multiphase microscale gas flow problems at a low computational cost.
 %It has been found that the MFS is an efficient tool for addressing two-dimensional multiphase microscale gas flow problems at a low computational cost.
Moreover, the findings also show that the CCR model solved with the method of fundamental solutions is able to describe rarefaction effects, like transpiration flows and thermal stress, generally well.  
\end{abstract}

\maketitle

\thispagestyle{empty}

\section{\label{sec:intro}Introduction}
The study of rarefied gases covers numerous applications, including flows caused by evaporation and condensation, upper-atmospheric dynamics, modeling of airborne particles, the reflective and reactive properties of gases interacting with solid and liquid surfaces and so on. Rarefied gas flows are characterized by a dimensionless parameter, the Knudsen number ($\mathrm{Kn}$), which is the ratio of the mean free path $\lambda$ of the gas and a characteristic length scale $L$ in the problem. 
For very small values of the Knudsen number ($\mathrm{Kn}\lesssim 0.01$), the classical continuum theories, namely the Euler and Navier--Stokes--Fourier (NSF) equations, are quite effective in capturing rarefaction effects but they fall short of doing so when the Knudsen number is not very small.
Although the NSF equations fail to capture several non-equilibrium phenomena (like non-homogeneity in pressure profile and unusual temperature dip in the Poiseuille flow \citep{OSA1989,TTS2009,RRPM2016}, heat flux direction opposite to the temperature gradient or the cross effects where heat flows from a low-temperature region to a high-temperature region \citep{ST2008,MRS2015,RTS2013}), yet, by exploiting the coupling among the thermodynamic forces and fluxes to form a closed system, the range of applicability of the new equations is enhanced in comparison to the NSF equations.
%The addition of second-order elements to the NSF equations and boundary conditions improves the modelling of moderate rarefaction effects.
%A model that exploits the coupling among the thermodynamic forces and fluxes has recently been propounded by Rana et al.~\citep{RGS2018}. which provides an improved set of constitutive relations called the coupled constitutive relations as they couple the stress and heat flux through a coupling coefficient. 
A model that exploits the coupling among the thermodynamic forces and fluxes to yield an improved set of constitutive relations for the stress and heat flux appearing in the conservation laws has recently been propounded by Rana \emph{et al.}~[\onlinecite{RGS2018}].
The constitutive relations for the stress and heat flux obtained in this model are coupled through a coupling coefficient; hence they are referred to as the coupled constitutive relations (CCR), and the model wherefore is referred to as the CCR model.
In the linearized and steady state, the CCR model reduces to the linearized Grad 13-moment (G13) equations \cite{Grad1949} in the steady state as a special case, and on taking the coupling coefficient as zero, the CCR model reduces to the original NSF equations.
Owing to its simplicity and viable features, the CCR model has been applied successfully to some problems pertaining to rarefied gas flows \cite{RSCLS2021, MS2022}. 
Although, there do exist other models, such as the regularized 13-moment (R13) equations \cite{ST2003, Struchtrup2005}, the regularized 26-moment (R26) equations \cite{GuEmerson2009}, etc., that can describe rarefied gas flows somewhat more accurately than the CCR model, especially for flows at moderate Knudsen numbers, we shall use the CCR model due to its simplicity in this work. 

In this paper, we shall focus our attention to exploring rarefied gas flow problems in (quasi) two dimensions (2D). 
The reason for this is twofold: firstly, for a symmetric uniform flow in three dimensions (3D), it is sufficient to study the problem in 2D, thanks to the symmetry along the transverse direction,
and secondly, there are some intriguing problems in 2D that do not arise in 3D, such as Stokes' paradox \citep{Lamb1932}, which states the non-existence of the steady-state solution to
Stokes' equation in 2D. 
Furthermore, we shall investigate
the problems numerically using a truly meshless numerical technique introduced by Kupradze and Aleksidze\citep{KA1964} known as the method of fundamental solutions (MFS).%

The MFS is a meshfree method that yields remarkably good results with a significantly less computational effort if the singularity points (also referred to as the source points) are placed at proper locations. 
The meshfree feature of the MFS is especially useful in the situations wherein changes in the shape of the domain are needed, e.g., in shape optimization and inverse problems.
This is because the MFS does not require creating a mesh over the entire domain, which itself could be a very time-consuming and computationally-expensive task depending on the complexity of the domain.
In the MFS, an approximate solution of a (linear) boundary value problem is expressed as a superposition of Green's functions, referred to as the fundamental solutions, and the %corresponding 
boundary conditions are satisfied at several locations on the boundary, referred to as the boundary nodes or collocation points, aiming to determine the unknown coefficients in the linear combination. %
%The MFS is a boundary-type meshfree approach in which an approximate solution of a (linear) boundary value problem is expressed as a linear combination of numerous singular functions, referred to as the fundamental solutions, and the %corresponding 
%boundary conditions are satisfied at several locations on the boundary, referred to as the boundary nodes or collocation points, aiming to determine the unknown coefficients in the linear combination. 
Apart from being time-efficient due to reduced spatial dimension in boundary discretization, the quality of being free from integrals makes the MFS peerless among other meshfree methods (such as the boundary element method \citep{brebbia1994BEM}, finite point method \citep{finite_pt_method}, diffuse element method \citep{diffuse_element}, element-free Galerkin method \citep{element_free_galerkin}) that involve complex integrals. 
The MFS has proven to be an efficient executable numerical scheme in various areas, such as thermoelasticity, electromagnetics, electrostatics, wave scattering, inverse problems and fluid flow problems; see, e.g.~Refs.~[\onlinecite{LFS2021, BK2001, FKM2003, KLM2011, YJFMT2006, LC2016}].
Moreover, the MFS is also suitable for the analysis of problems involving shape optimization, moving boundary and unknown boundary \citep{FR2002, YJFMT2006, chen2008, SLL2012, Alves2009}, since the problems of modeling and satisfying boundary conditions are relatively simple for them.
All these advantages make the application of the MFS to the CCR model evidently favorable.
Several researchers have employed the MFS to solve the Helmholtz-, harmonic- and biharmonic-type boundary value problems in 2D as well as in 3D, see e.g., Refs.~[\onlinecite{PKG1998, ML2005}].
The MFS works as a good numerical strategy if the fundamental solutions to the problem are predefined. 
In the past few years, there has been a surge of interest in employing the MFS to various models for rarefied gas flows, for instance to the NSF, G13, R13 and CCR models~\citep{LC2016, CSRSL2017, RSCLS2021}, because the predefined fundamental solutions of the well-known equations, such as the Laplace, Helmholtz and biharmonic equations, can be exploited to determine the fundamental solutions for the NSF, G13, R13 and CCR models. 
Nevertheless, all the works on the MFS for rarefied gas flows have investigated the  problems in 3D only.
%
%Several works have been carried out to derive the fundamental solutions and to implement the MFS in 3D for studying rarefied gases using different models, such as the NSF, Grad 13-moment, regularized 13-moment and CCR models~\citep{LC2016,CSRSL2017,RSCLS2021}.
But, for quasi two-dimensional flow problems, it is not really necessary to solve the full three-dimensional problem as the flow profiles obtained in a cross section perpendicular to the transverse direction remain the same in any cross section perpendicular to the transverse direction.
Thus, a quasi two-dimensional study of a full three-dimensional problem (where one dimension in the problem is much larger than the other two) reduces the computational cost of any numerical technique exceedingly.
Interestingly, in the case the MFS, the two-dimensional fundamental solutions for a model cannot be deduced directly from its three-dimensional counterpart due to the fact that the associated Green's functions are entirely different in
2D and 3D. 
%It is, however, important to note that the two-dimensional fundamental solutions for a model cannot be deduced directly from its three-dimensional counterpart due to the fact that the associated Green’s functions are entirely different in 2D and 3D.
% The 2D fundamental solutions have a logarithmic term causing the non-existence of a bounded steady-state solution for Stokes' equations in 2D. This phenomenon referred to as Stokes' paradox is not seen in the 3D case. The relevance for Stokes' paradox with the 2D-CCR model is beyond the scope of this paper and will be considered as a future purview.
%{\color{red}The main advantage of using the MFS in 2D is the notable reduction in computational cost compared to the MFS in 3D, when the flow is quasi two-dimensional. This is due to the fact that the number of boundary nodes and singularity points increases crucially while implementing the MFS in 3D for quasi two-dimensional flows. 
%Moreover, the representation of complex geometries becomes simpler for the MFS in 2D as it only necessitates two-dimensional coordinates, simplifying the setup and reducing computational effort in terms of boundary discretization. 
%Additionally, the visualization and analysis of results are enhanced in 2D as the outcomes represented on a plane facilitate the interpretation and comprehension of the physical phenomena being studied.}
Therefore the main objectives of the paper are (i) to determine the two-dimensional fundamental solutions of the linearized CCR model
% which is a somewhat better model than the NSF equations to study moderately rarefied gases and is simpler as compared to other kinetic models that are used to study microscale flows.
% These fundamental solutions are derived by means of predefined fundamental solutions of some well-known equations like the Laplace and biharmonic equations. 
and (ii) to implement the determined fundamental solutions in a numerical framework. 
To gauge the accuracy of the developed numerical framework, the obtained numerical results are also validated against those obtained with other models for a few problems existing in the literature.
%The three dimensional MFS has been done with the CCR model in \citep{RSCLS2021} after extension of fundamental solutions obtained by \citep{LC2016} which describes the MFS with NSF, G13 as well as R13 closure named as Stokeslet, Gradlet and Reglet respectively. We primarily focus on the MFS in two dimensions because the fundamental solutions are different in 2D and 3D due to which nature of overall numerical solution changes.
%For validation, we 
We pick two internal-flow problems from Refs.~[\onlinecite{Onishi1977, Aoki_Sone_1989}] that have been investigated in these references with the linearized Bhatnagar--Gross--Krook (BGK) model \citep{BGK1954} [also referred to as the Boltzmann--Krook--Welander (BKW) kinetic model by some authors \cite{Onishi1977, Aoki_Sone_1989, Sone2007}].
In the first problem, the evaporation and condensation of a mildly rarefied vapor confined between two coaxial cylinders is studied while in the second problem,  a temperature-driven rarefied gas flow between two non-coaxial cylinders is investigated.

In the MFS, positioning of the singularity points has been a widely-discussed issue in order to achieve accurate results \citep{Alves2009, CKL2016, WLQ2018, CH2020}
due to the fact that the linear system resulting from the MFS can have an ill-conditioned coefficient matrix \citep{Alves2009}, and there is a trade-off between the accuracy and well conditioning.
%This is because the MFS typically leads to the associated linear system having an ill-conditioned matrix \citep{Alves2009} and there is a trade-off between the accuracy and well conditioning. 
For meshfree methods, including the MFS, \citeauthor{Alves2009} [\onlinecite{Alves2009}] states, ``In these methods a sort of uncertainty principle occurs---\emph{we cannot get both accurate results and good conditioning}---one of the two is lost.''
In this paper, we also demonstrate a method to determine an appropriate location of singularity points for obtaining the solutions of a desired accuracy using an approach based on the effective condition number as discussed in Refs.~[\onlinecite{DML2009, CNYC2023, WL2011}].
%
%Another shortcoming of the MFS comes up with the formation of ill-conditioned collocation matrix in some problems \citep{CH2020, DML2009, CNYC2023, WL2011}. 
%Therefore, we also investigate the satisfactory location of singularities using an approach based on the effective condition number as discussed in Refs.~[\onlinecite{DML2009, CNYC2023, WL2011}].
%
%A well-known shortcoming of the MFS is that it \textcolor{blue}{might} yield the results that are highly sensitive toward the location of the singularity points (also referred to as the source points), and that a high accuracy \textcolor{blue}{can be} accompanied with an ill-conditioned collocation matrix \citep{Alves2009, CKL2016, WLQ2018, CH2020}. 
%Therefore, we also investigate the optimal location of singularities using an approach based on the effective condition number as discussed in Refs.~[\onlinecite{DML2009, CNYC2023, WL2011}].
% ---one is to study flow between two coaxial evaporating cylinders as done in \citep{Onishi1977}. This problem includes the phase-transitional flow simulation which establishes the validity of the 2D MFS for multiphase problems. The second is done by temperature-driven flow problem between two non-coaxial cylinders as discussed in \citep{Aoki_Sone_1989}. 

The remainder of the paper is structured as follows. 
The linearized CCR model and the generalized boundary conditions associated with it are outlined in Sec.\,\ref{sec:balance_eqs}. 
The two-dimensional fundamental solutions for the CCR model are determined in Sec.\,\ref{sec:solutions}. 
The technique to apply the MFS by forming a system of equations for any arbitrary geometry is discussed in Sec.\,\ref{sec:system}. 
The implementation of the MFS along with its validation (i) for the problem of a vapor flow between two coaxial cylinders is  demonstrated in Sec.\,\ref{sec:prob1}
and (ii) for the problem of a temperature-driven rarefied gas flow between two non-coaxial cylinders is discussed in Sec.\,\ref{sec:prob2}. 
The location of singularities based on the effective condition number approach is examined in Sec.\,\ref{sec:location}. 
The paper ends with conclusions and outlook in Sec.\,\ref{sec:conclusion}.
%The linearized conservation laws closed with the help of the CCR closure and the generalized boundary conditions derived in [\onlinecite{RSCLS2021}] has been explained in Sec.\,\ref{sec:balance_eqs}. The fundamental solutions for the set of equations are derived in Sec.\,\ref{sec:solutions} and the technique to apply the MFS by forming a system of equations for any arbitrary geometry is discussed in Sec.\,\ref{sec:system}. The next Sec.\,\ref{sec:prob1} demonstrates the implementation and validation of the MFS for the vapor flow between two coaxial cylinders. The second problem of temperature-driven flow between two non-coaxial cylinders is discussed in Sec.\,\ref{sec:prob2}. The location of singularities based on the effective condition number approach has been examined in Sec.\,\ref{sec:location}. The paper ends with conclusions and outlook in Sec.\,\ref{sec:conclusion}.
\section{\label{sec:balance_eqs}The linearized CCR model and boundary conditions}
The CCR model consists of the conservation laws---the balance equations for the mass, momentum and energy---closed with the constitutive relations for the stress and heat flux, which are coupled with each other through a coupling coefficient.
The full details of the CCR model can be found in Ref.~[\onlinecite{RGS2018}].
In this work, we require them in the linearized form.
To this end, we convert the mass, momentum and energy balance equations and the coupled constitutive relations into a linear-dimensionless form by assuming small perturbations in flow fields from their respective equilibrium values. 
The velocity, stress and heat flux in the equilibrium state vanish whereas the density and temperature in the equilibrium state are constants $\tilde{\rho}_0$ and $\tilde{T}_0$, respectively. 
The dimensionless perturbations in the density $\tilde{\rho}$ and temperature $\tilde{T}$ from their values in the equilibrium are given by 
\begin{align}
\label{scaledrhoT}
\rho=\frac{\tilde{\rho}-\tilde{\rho}_0}{\tilde{\rho}_0}
\quad\text{and}\quad
T=\frac{\tilde{T}-\tilde{T}_0}{\tilde{T}_0},
\end{align}
respectively. 
Similarly, the dimensionless perturbations in the velocity $\tilde{\bm{v}}$, stress tensor $\tilde{\bm{\sigma}}$ and heat flux $\tilde{\bm{q}}$ from their values in the equilibrium are given by
\begin{align}
\label{scalednoneqbquan}
\bm{v}=\frac{\tilde{\bm{v}}}{\sqrt{\tilde{\theta}_0}},
\quad 
\bm{\sigma}=\frac{\tilde{\bm{\sigma}}}{\tilde{\rho}_0 \tilde{\theta}_0}
\quad\text{and}\quad
\bm{q}=\frac{\tilde{\bm{q}}}{\tilde{\rho}_0 (\tilde{\theta}_0)^{3/2}}, 
\end{align}
respectively, where $\tilde{\theta}_0 = \tilde{R}\tilde{T_0}$ with $\tilde{R}$ being the gas constant. 
The linearized equation of state $p\approx\rho+T$ gives the dimensionless perturbation in the pressure from its equilibrium value $\tilde{p}_0=\tilde{\rho}_0\tilde{\theta}_0$. 
For the sake of simplicity, the field variables with tilde are the quantities with dimensions while those without tilde are the dimensionless quantities throughout the paper.

Considering $\tilde{L}$ to be the characteristic length scale, the dimensionless position vector is $\bm{r}=\tilde{\bm{r}}/\tilde{L}$. 
Inserting these dimensionless variables into the CCR model \cite{RGS2018} and dropping all nonlinear terms in the perturbed variables, one readily obtains the linear-dimensionless CCR model. 
Here, we present them directly.
The linear-dimensionless mass, momentum and energy balance equations in the steady state read
\begin{align}
\label{mass_bal}
\bm{\nabla}\cdot\bm{v}=0,
\\
\label{mom_bal}
\bm{\nabla} p+\bm{\nabla}\cdot\bm{\sigma}=\bm{0},
\\
\label{energy_bal}
\bm{\nabla}\cdot\bm{q}=0,
\end{align}
%The conservation laws (\ref{mass_bal})--(\ref{energy_bal}) are not closed as they contain stress tensor $\bm{\sigma}$ and heat flux $\bm{q}$, which are unknowns. To close the system, we adopt linearized coupled constitutive relations \citep{RGS2018}
and, to close the system \eqref{mass_bal}--\eqref{energy_bal}, we adopt the linearized coupled constitutive relations\citep{RGS2018}
\begin{align}
\label{ccr_1}
\bm{\sigma}&=-2\mathrm{Kn}\overline{\bm{\nabla} \bm{v}}-2\alpha_0 \mathrm{Kn}\overline{\bm{\nabla} \bm{q}},
\\
\label{ccr_2}
\bm{q}&=-\frac{c_p \mathrm{Kn}}{\mathrm{Pr}}(\bm{\nabla} T+\alpha_0\bm{\nabla}\cdot\bm{\sigma}),
\end{align}
where $\alpha_0$ is the coupling coefficient through which constitutive relations \eqref{ccr_1} and \eqref{ccr_2} are coupled; 
$c_p = \tilde{c}_p / \tilde{R}$ with $\tilde{c}_p$ being the specific heat capacity of the gas at a constant pressure; and
\begin{align}
\mathrm{Pr}=\frac{\tilde{c}_p\tilde{\mu}_0}{\tilde{\kappa}_0}
\quad\text{and}\quad
\mathrm{Kn}=\frac{\tilde{\mu}_0}{\tilde{\rho}_0\sqrt{\tilde{\theta}_0}\tilde{L}}
\end{align}
are the Prandtl number and Knudsen number, respectively, with $\tilde{\mu}_0$ and $\tilde{\kappa}_0$ being the viscosity and thermal conductivity at the equilibrium state. 
The quantities $\overline{\bm{\nabla} \bm{v}}$ and $\overline{\bm{\nabla} \bm{q}}$ in \eqref{ccr_1} are the symmetric-tracefree parts of the tensors $\bm{\nabla} \bm{v}$ and $\bm{\nabla} \bm{q}$, respectively. 
For a $d$-dimensional vector $\bm{\psi}$, the symmetric-tracefree part of the tensor $\bm{\nabla} \bm{\psi}$ is defined as \cite{Gupta2020}
\begin{align}
\label{stf_tensor}
\overline{\bm{\nabla} \bm{\psi}}&=\frac{1}{2}\Big[{\bm{\nabla} \bm{\psi}}+(\bm{\nabla} \bm{\psi})^\mathsf{T}\Big]-\frac{1}{d}(\bm{\nabla}\cdot\bm{\psi})\bm{I},
\end{align}
where $\bm{I}$ is the identity tensor in $d$ dimensions.
For three-dimensional and quasi two-dimensional problems, $d=3$.
%Since we shall be dealing with the problems in 2D, $d=2$ is fixed throughout the paper.
%
Furthermore, the dimensionless specific heat of a gas at a constant pressure is $c_p = (5+\mathsf{n})/2$, where $\mathsf{n}$ is a positive number that accounts for the internal degrees of freedom in a polyatomic gas.
%
%Furthermore, $c_p = \tilde{c}_p/\tilde{R}$ in \eqref{ccr_2} is the dimensionless specific heat of the gas (which could be monatomic, diatomic or polyatomic) and is given by $(5+\mathsf{n})/2$, where $\mathsf{n}$ is a positive integer that accounts for the rotational degrees of freedom in a polyatomic gas. 
For monatomic gases, there is no rotational degree of freedom; consequently, $\mathsf{n}=0$ and $c_p=5/2$ for monatomic gases. 
We shall only deal with monatomic gases in this paper, and hence $c_p=5/2$ throughout this paper. 
Equations \eqref{mass_bal}--\eqref{energy_bal} closed with \eqref{ccr_1} and \eqref{ccr_2} are referred to as the linearized CCR model.
%{\color{red}The first term on
%the right-hand side of the constitutive relation \eqref{ccr_1} is the viscous stress term, which represents the Navier--Stokes contribution to the stress while the second term is the thermal stress term rendered by the heat flux \citep{MRS2015, Sone2007}. 
%The first term on the right-hand side of the constitutive relation \eqref{ccr_2} represents the Fourier contribution to the heat flux \st{while the second term accounts for the anti-Fourier effects in rarefied gases}\citep{MRS2015}.}
For $\alpha_0=0$, the linearized CCR model reduces to the linearized NSF equations and for $\alpha_0=2/5$, the steady-state linearized CCR model reduces to the steady-state linearized G13 equations. 
The parameter $\alpha_0$ in the case of the CCR model is taken as $0.3197$, the value of $\alpha_0$ for hard sphere molecules \cite{RGS2018}, throughout this paper.
Since we shall be comparing the results obtained in the present work with those obtained with the BGK model, for which the Prandtl number is unity \cite{Struchtrup2005}, $\mathrm{Pr}=1$  throughout this paper.
%Also, the parameter $\alpha_0$ is taken as $0.3197$, the value of $\alpha_0$ for hard sphere molecules \cite{RGS2018}, {\color{blue}in the case of the CCR model}.
%
%Also, the value of the coefficient $\alpha_0$ for hard-sphere molecules is $0.3197$~[\onlinecite{RGS2018}], which has been used throughout this paper unless mentioned otherwise. 
%In relations (\ref{ccr_1}) and (\ref{ccr_2}), $\alpha_0$ is the coupling coefficient such that for $\alpha_0=0$ and $2/5$, the linear CCR equations reduce to the linearized NSF and Grad $13$-moment equations, respectively; for more details, see [\onlinecite{RGS2018}] and [\onlinecite{RSCLS2021}].
%Further,
%\begin{align*}
%\overline{\bm{\nabla} \bm{v}}&=\frac{1}{2}({\bm{\nabla} \bm{v}}+(\bm{\nabla} \bm{v})^\mathsf{T})-\frac{1}{d}\bm{I}(\bm{\nabla}\cdot\bm{v}),
%\\ \overline{\bm{\nabla} \bm{q}}&=\frac{1}{2}({\bm{\nabla} \bm{q}}+(\bm{\nabla} \bm{q})^\mathsf{T})-\frac{1}{d}\bm{I}(\bm{\nabla}\cdot\bm{q})
%\end{align*}
%represent the symmetric, traceless component of rank-$2$ tensors $\bm{\nabla} \bm{v}$ and $\bm{\nabla} \bm{q}$ in $d$-dimensions respectively, and $\bm{I}$ is identity tensor. 

For quasi two-dimensional flows, let us say in the $x_1 x_2$-plane, the field variables do not change in the direction perpendicular to the plane of the flow, i.e.~they do not change along the  $x_3$-direction.
As a result, the CCR model [Eqs.~\eqref{mass_bal}--\eqref{ccr_2}] for a quasi two-dimensional flow in the $x_1 x_2$-plane reduces to
\begin{align}
\label{mass_bl_q2D}
\frac{\partial v_i}{\partial x_i} &= 0,
\\
\label{mom_bl_q2D}
\frac{\partial p}{\partial x_i}+\frac{\partial \sigma_{ij}}{\partial x_j} &= 0,
\\
\label{energy_bl_q2D}
\frac{\partial q_i}{\partial x_i} &= 0,
\end{align}
\begin{align}
\label{ccr_1_q2D}
\sigma_{ij}=&-2\mathrm{Kn}\left[\frac{1}{2}\left(\frac{\partial v_i}{\partial x_j}+\frac{\partial v_j}{\partial x_i}\right) - \frac{1}{3} \delta_{ij} \frac{\partial v_\ell}{\partial x_\ell}\right]
\nonumber\\
&-2\alpha_0\mathrm{Kn}\left[\frac{1}{2}\left(\frac{\partial q_i}{\partial x_j}+\frac{\partial q_j}{\partial x_i}\right) - \frac{1}{3} \delta_{ij} \frac{\partial q_\ell}{\partial x_\ell}\right],
\\
\label{ccr_2_q2D}
q_i=&-\frac{c_p \mathrm{Kn}}{\mathrm{Pr}}\left(\frac{\partial T}{\partial x_i}+\alpha_0 \frac{\partial \sigma_{ij}}{\partial x_j}\right),
\end{align}
where the indices $i,j$ and $\ell$ can take the values $1$ and $2$ only, $\delta_{ij}$ is the Kronecker delta and the Einstein summation applies over the repeated indices in a term.
It may be noted that Eq.~\eqref{mom_bl_q2D} represents two equations: for $i=1$ the momentum balance equation in the $x_1$-direction and for $i=2$ the momentum balance equation in the $x_2$-direction, and that the momentum balance equation in the $x_3$-direction is identically satisfied.
It is also worthwhile noting that $\sigma_{11} + \sigma_{22} = 0$ in view of Eqs.~\eqref{mass_bl_q2D} and \eqref{energy_bl_q2D}, which is consistent with the fact that the stress tensor $\bm{\sigma}$ is tracefree because $\sigma_{33} = 0$ for quasi two-dimensional flows in the $x_1 x_2$-plane.
%
%\begin{align}
%\frac{\partial v_1}{\partial x} + \frac{\partial v_2}{\partial y} &= 0,
%\\
%\frac{\partial p}{\partial x} 
%+ \frac{\partial \sigma_{11}}{\partial x}
%+ \frac{\partial \sigma_{12}}{\partial y} &= 0,
%\\
%\frac{\partial p}{\partial y} 
%+ \frac{\partial \sigma_{12}}{\partial x}
%+ \frac{\partial \sigma_{22}}{\partial y} &= 0,
%\\
%\frac{\partial q_1}{\partial x} + \frac{\partial q_2}{\partial y} &= 0,
%\end{align}
Thus, the CCR model for a quasi two-dimensional flow in the $x_1 x_2$-plane [Eqs.~\eqref{mass_bl_q2D} and \eqref{ccr_2_q2D}] essentially consists of the unknown field variables $v_1, v_2, p, T, \sigma_{11}, \sigma_{12}, q_1, q_2$.  
Therefore, the fundamental solutions of the CCR model for a quasi two-dimensional flow---determined in the next section---will be referred to as the two-dimensional fundamental solutions of the CCR model or the fundamental solutions of the CCR model in 2D.

%\textcolor{blue}{The flow problems investigated in the current work are deemed to be quasi two-dimensional, indicating that the flow is mainly confined to the $x$-$y$ plane and the $z$ direction has a scale that is much greater than the other two directions. For instance, the velocity component in the $z$ direction vanishes and we can express the dimensionless velocity vector as $\bm{v}=(v_1(x,y),v_2(x,y),0)^\mathsf{T}$, where $v_1$ and $v_2$ are the corresponding velocity components in the $x$ and $y$ directions.}

The thermodynamically-consistent boundary conditions complementing the linear CCR model have been derived in Ref.~[\onlinecite{RSCLS2021}]. 
For a three-dimensional problem, the boundary conditions complementing the linear CCR model are given in Eqs.~$(4.2a)$, $(4.2b)$, $(4.3a)$ and $(4.3b)$ of Ref.~[\onlinecite{RSCLS2021}].
Equations $(4.2a)$ and $(4.2b)$ of Ref.~[\onlinecite{RSCLS2021}] are the boundary conditions on the normal components of the mass and heat fluxes, respectively, while Eqs.~$(4.3a)$ and $(4.3b)$ of Ref.~[\onlinecite{RSCLS2021}] are the boundary conditions on the shear stress---two conditions due to two tangential directions in 3D.
Since for a quasi two-dimensional flow in the $x_1 x_2$-plane, the wall normal direction and one tangential direction are in the $x_1 x_2$-plane while the other tangential direction is along the $x_3$-direction,
%Since in 2D, there will be only one tangential direction, there will be only one tangential direction, 
boundary condition $(4.3b)$ of Ref.~[\onlinecite{RSCLS2021}] is irrelevant in the present work and the superscript `$(1)$' can be dropped from the unit tangent vector $\bm{t}^{(1)}$ in $(4.3a)$ of Ref.~[\onlinecite{RSCLS2021}] for simplicity.
Thus, the linear-dimensionless boundary conditions complementing the linearized CCR model for a quasi two-dimensional flow read \cite{RSCLS2021}
%
%The first two boundary conditions are the equations $(4.2a)$ and $(4.2b)$ in [\onlinecite{RSCLS2021}] and since there is only one tangential direction in 2D, the third boundary condition for the CCR model is obtained by omitting equation $(4.3b)$ and replacing $\bm{t}^{(1)}$ with $\bm{t}$ in equation $(4.3a)$. The boundary conditions for the present case read
%We adopt the boundary conditions derived 
%in \citep{RSCLS2021} for the CCR model in 3D which can be applied to the 2D-CCR model if we consider one tangent at a point in 2D. The boundary conditions read
%Since walls of cylinders are considered to be condensing, we use boundary conditions to observe velocity and heat flux due to temperature and pressure difference at interface. Also we consider the flow induced by tangential heat flux. These are taken from \citep{RSCLS2021}
\begin{align}
\label{bc_1}
(\bm{v}-\bm{v}^I)\cdot \bm{n}=& -\eta_{11}(p-p_{\mathrm{sat}} + \bm{n}\cdot\bm{\sigma}\cdot\bm{n})
\nonumber\\
&+\eta_{12}(T-T^I+\alpha_0 \bm{n}\cdot\bm{\sigma}\cdot\bm{n}),
\end{align}
\begin{align}
\label{bc_2}
\bm{q}\cdot \bm{n} =& \medspace\eta_{12}(p-p_{\mathrm{sat}} + \bm{n}\cdot\bm{\sigma}\cdot\bm{n})
\nonumber\\
&-(\eta_{22}+2\tau_0)(T-T^I+\alpha_0 \bm{n}\cdot\bm{\sigma}\cdot\bm{n}),
\end{align}
\begin{align}
\label{bc_3}
\bm{t}\cdot\bm{\sigma}\cdot\bm{n} =& -\varsigma (\bm{v}-\bm{v}^I+\alpha_0 \bm{q})\cdot\bm{t},
\end{align}
where $\bm{n}$ and $\bm{t}$ are the unit normal and tangent vectors, respectively. 
In boundary conditions \eqref{bc_1}--\eqref{bc_3}, $\eta_{ij}$'s,  for $i,j\in\{1,2\}$  are the Onsager reciprocity coefficients, which from Sone's asymptotic kinetic theory \citep{Sone2007} turn out to be
\begin{align}
\left.
\begin{aligned}
\label{coff}
\eta_{11}&=0.9134\sqrt{\frac{2}{\pi}}\frac{\vartheta}{2-\vartheta},
\\
\eta_{12}&=0.3915\sqrt{\frac{2}{\pi}}\frac{\vartheta}{2-\vartheta},
\\
%\text{and}\quad
\eta_{22}&=0.1678\sqrt{\frac{2}{\pi}}\frac{\vartheta}{2-\vartheta}
\end{aligned}
\right\}
\end{align}
under the assumption of the accommodation coefficient being unity (which also holds true for the diffuse reflection boundary condition).
The parameter $\vartheta$ in the above coefficients is the evaporation/condensation coefficient. 
%with $\vartheta$ being evaporation/condensation coefficient.
%which is independent of the impact energy of molecules. 
For canonical boundaries and phase-change boundaries, $\vartheta=0$ and $1$, respectively, are the largely accepted values of $\vartheta$ in the literature. 
The temperature-jump and velocity-slip coefficients are given by \cite{RSCLS2021}
\begin{align}
\tau_0=0.8503\sqrt{\frac{2}{\pi}}\quad\text{and}\quad 
\varsigma = 0.8798\sqrt{\frac{2}{\pi}},
\end{align} 
respectively. 
Furthermore,  $\bm{v}^I$, $T^I$ and $p_{\mathrm{sat}}$ in boundary conditions \eqref{bc_1}--\eqref{bc_3} represent the velocity, temperature and saturation pressure at the interface.
It is important to note that the coefficients $\alpha_0$ in boundary conditions \eqref{bc_1}--\eqref{bc_3} are actually the fitting parameters and could be different from the coupling coefficient $\alpha_0$. 
Moreover, the coefficient $\alpha_0$ in each of boundary conditions \eqref{bc_1}--\eqref{bc_3} could also be different from each other.
The only reason that the coefficients $\alpha_0$ in boundary conditions \eqref{bc_1}--\eqref{bc_3} have been taken as the same as the coupling coefficient in the CCR model because the boundary conditions obtained in this way are thermodynamically consistent \cite{RGS2018}.

% Also, walls are considered to be stationary in the problems  for application   i.e. $\bm{v}^I=0$.
%The system of equations (\ref{mass_bal})--(\ref{ccr_1}) is solved with the help of boundary conditions (\ref{bc_1})--(\ref{bc_3}).
%
\section{\label{sec:solutions} Derivation of the fundamental solutions of the CCR model}
The fundamental solutions of the CCR model in 3D have already been derived in Ref.~[\onlinecite{RSCLS2021}].  
However,  as mentioned in Sec.\,\ref{sec:intro},  the fundamental solutions of a model in 2D and 3D are independent of each other because the inherent Green's functions are independent of each other in 2D and 3D;
consequently, the two-dimensional fundamental solution of a model cannot be determined from its three-dimensional counterpart in general. 
Therefore,  we derive the  fundamental solutions of the CCR model in 2D from scratch in this section.
%We focus on two-dimensional fundamental solutions in this paper which are independent of the fundamental solutions in 3D as the Green's functions in 2D and 3D are entirely different. 
%To derive the fundamental solutions to the CCR model in 2D,  
To this end,  we add a Dirac delta forcing term of strength $f_i$ ($i\in\{1,2\}$)
%$\bm{f}$ 
on the right-hand side of the momentum balance equation to represent a (vector) point force, and a point heat source of strength $g$  on the right-hand side of the energy balance equation.  
Furthermore,  to deal with phase-change effects at the liquid-vapor interface,  a point mass source of strength $h$ is also added  on the right-hand side of the mass balance equation.  
For determining the fundamental solutions of a system of partial differential equations, it is customary to consider only one point source  at a time and then to superimpose the solutions obtained by taking each point source at a time in order to incorporate the effects of all point sources; see,  e.g., Refs.~[\onlinecite{LC2016, RSCLS2021}].
%The fundamental solutions are usually derived by taking one source strength at a time and then the superposition of the solutions is employed to incorporate the effects of all point sources; see,  e.g. ~Refs.~ [\onlinecite{LC2016, RSCLS2021}].
Nevertheless,  we take all three point sources $\bm{f}\equiv (f_1,f_2)^{\mathsf{T}}$, $g$ and $h$ simultaneously and solve the resulting system of equations altogether.  
We have verified---although not shown here for brevity---that this procedure also yields exactly the same solution as that obtained  by superimposing the solutions obtained by solving the systems separately with one point source at a time.
%
%and solve the system in tensorial notations for the sake of simplicity.  
%It turns out (although not shown here for brevity) that both techniques yield the same fundamental solutions.
% Both the above approaches to derive the fundamental solutions are equivalent.
%In the literature(?), 2D and 3D stokeslets are well-known for studying the flow response to point body forces. However, none of these result in a mass flux. Instead of utilising stokeslet as a source point, we use multiple strengths as source points to find fundamental solutions for the CCR model. We add a mass source of strength h to the mass balance equation to examine phase change effects at the liquid-vapor interface. A similar Sourcelet solution has been derived in \citep{RSCLS2021} for three dimensional phase change interface.
%
%It is easy to keep track of the derivation of the fundamental solutions of the CCR equations in the indicial notation. 
%Therefore,  we shall derive the fundamental solutions of the CCR equations first in the indicial notation and then express them in the vectorial/tensorial notation.
%For that,  let us first write down the linearized CCR model (with the point source terms) in the indicial notation.
%The mass, momentum and energy balance equations \eqref{mass_bal}--\eqref{energy_bal}---with the point source terms---in the indicial notation read

To determine the fundamental solutions of the CCR model in 2D, the mass, momentum and energy balance equations \eqref{mass_bl_q2D}--\eqref{energy_bl_q2D} are written with the point source terms on their right-hand sides. These equations read
\begingroup
\allowdisplaybreaks
\begin{align}
\label{mass_bl}
\frac{\partial v_i}{\partial x_i}&=h\,\delta(\bm{r}),
\\
\label{mom_bl}
\frac{\partial p}{\partial x_i}+\frac{\partial \sigma_{ij}}{\partial x_j}&=f_i \,\delta(\bm{r}),
\\
\label{energy_bl}
\frac{\partial q_i}{\partial x_i}&=g\,\delta(\bm{r}),
\end{align}
\endgroup
where $\bm{r}= (x_1,x_2)^{\mathsf{T}}$.
Equations \eqref{mass_bl}--\eqref{energy_bl} are closed with the CCR \eqref{ccr_1_q2D} and \eqref{ccr_2_q2D}.
We solve the system of Eqs.~\eqref{mass_bl}--\eqref{energy_bl}, \eqref{ccr_1_q2D} and \eqref{ccr_2_q2D} using the Fourier transformation.
For this, we define the Fourier transform pair (the Fourier transform and the inverse Fourier transform) as
\begin{align}
\label{ft}
\mathcal{F}\big(F(\bm{r})\big) 
=\hat{F}(\bm{k}) := \int_{\mathbb{R}^2} F(\bm{r}) \, \mathrm{e}^{\mathbbm{i} \, \bm{k} \cdot \bm{r}} \, \mathrm{d} \bm{r}
\end{align}
and
\begin{align}
\label{ftinv}
\mathcal{F}^{-1}\big(\hat{F}(\bm{k})\big) 
=F(\bm{r})
:=\frac{1}{(2\pi)^2}\displaystyle{ \int_{\mathbb{R}^2}}
\hat{F}(\bm{k}) \, \mathrm{e}^{-\mathbbm{i}\,\bm{k} \cdot \bm{r}} \, \mathrm{d}\bm{k},
\end{align} 
respectively.
Here, $\mathbbm{i}$ is the imaginary unit and $\bm{k}$ is the wavevector in the spatial-frequency domain.

Applying the Fourier transformation in Eqs.~\eqref{mass_bl}--\eqref{energy_bl}, \eqref{ccr_1_q2D} and \eqref{ccr_2_q2D} and using the fact that $\mathcal{F}[\delta(\bm{r})]=1$,  we obtain ($i,j,\ell \in \{1,2\}$)
\begin{align}
\label{mass_ft}
k_i \hat{v}_i &=\mathbf{\mathbbm{i}} \,h,
\\
\label{mom_ft}
k_i \hat{p} + k_j \hat{\sigma}_{ij} &= \mathbbm{i} \, f_i,
\\
\label{energy_ft}
k_i \hat{q}_i &= \mathbbm{i} \, g,
\end{align}
\begin{align}
\label{ccr1_ft}
\hat{\sigma}_{ij}=&\medspace \mathbbm{i} \,\mathrm{Kn} \bigg[ k_j(\hat{v}_i + \alpha_0\hat{q}_i)+k_i(\hat{v}_j+\alpha_0\hat{q}_j)\nonumber
\\
&-\frac{2}{3} \delta_{ij} k_\ell (\hat{v}_\ell+\alpha_0\hat{q}_\ell)\bigg],
\\[2ex]
\label{ccr2_ft}
\hat{q}_i =&\medspace \mathbbm{i}\frac{c_p \mathrm{Kn}}{\mathrm{Pr}}\left(k_i \hat{T}+\alpha_0 k_j\hat{\sigma}_{ij}\right),
\end{align}
where the variables with hat are the Fourier transforms of the corresponding field variables. 
Using Eqs.~\eqref{mass_ft} and \eqref{energy_ft},  Eq.~\eqref{ccr1_ft} simplifies to
\begin{align}
\label{ss1}
\hat{\sigma}_{ij}=& \medspace\mathbbm{i} \,\mathrm{Kn} \big[ k_j(\hat{v}_i + \alpha_0\hat{q}_i)+k_i(\hat{v}_j+\alpha_0\hat{q}_j)\big]
\nonumber\\ 
&+\frac{2}{3} \delta_{ij} \mathrm{Kn}(h+\alpha_0g).
\end{align}
Multiplying the above equation with  $k_j$ and $k_i k_j$, we obtain \begin{align}
\label{ss11}
k_j\hat{\sigma}_{ij}&=\mathbbm{i} \, \mathrm{Kn}\, k^2(\hat{v}_i+\alpha_0\hat{q}_i) - \frac{1}{3}\mathrm{Kn}\, k_i(h+\alpha_0 g),
\\
\label{ss12}
k_i k_j\hat{\sigma}_{ij}&=-\frac{4}{3}\mathrm{Kn}\, k^2(h+\alpha_0 g),
\end{align}
respectively, where $k_i k_i=|k_i|^2=k^2$ has been used.
%Multiplying $\omega_i$ in \eqref{ccr2_ft},
%\begin{align}
%\label{ss2}
%\omega_i \hat{q}_i&=\mathbbm{i}\frac{c_p \mathrm{Kn}}{\mathrm{Pr}}\left(\omega^2 \hat{T}+\alpha_0 \omega_i\omega_j\hat{ \sigma}_{ij}\right),
%\end{align}
Multiplying Eq.~\eqref{ccr2_ft} with $k_i$ and exploiting Eqs.~\eqref{energy_ft} and \eqref{ss12},  we obtain 
\begin{align}
\label{theta_ft}
\hat{T}=\frac{\mathrm{Pr}}{c_p \mathrm{Kn}} \frac{g}{k^2} + \frac{4}{3} \alpha_0 \mathrm{Kn}(h+\alpha_0 g).
\end{align}
Again, multiplying Eq.~\eqref{mom_ft} with $k_i$ and exploiting Eq.~\eqref{ss12},  we obtain 
\begin{align}
%\label{ss4}
%k^2\hat{p}+k_ik_j\hat{\sigma}_{ij}&=\mathbbm{i}k_i f_i
%\\
%\implies
\label{p_ft}
\hat{p}&=\mathbbm{i}\frac{k_i f_i}{k^2} + \frac{4}{3}\mathrm{Kn}(h+\alpha_0 g).
\end{align}
Now, from Eqs.~\eqref{mom_ft} and \eqref{p_ft}, one can easily write \begin{align}
\label{ss5}
k_j\hat{\sigma}_{ij}&=\mathbbm{i} f_i-\mathbbm{i}\frac{k_i k_j f_j}{k^2}-\frac{4}{3}k_i \mathrm{Kn}(h+\alpha_0 g).
\end{align}
Substituting the value of $\hat{T}$ from Eq.~\eqref{theta_ft} and the value of $k_j\hat{\sigma}_{ij}$ from Eq.~\eqref{ss5} into Eq.~\eqref{ccr2_ft}, we obtain
\begin{align}
\label{q_ft}
\hat{q}_i&=\mathbbm{i}\frac{k_i g}{k^2} -\frac{c_p \mathrm{Kn}}{\mathrm{Pr}} \alpha_0 f_j\left(\delta_{ij}-\frac{k_ik_j}{k^2}\right).
\end{align}
Now, from Eqs.~\eqref{ss11}, \eqref{ss5} and \eqref{q_ft},
%On equating \eqref{ss11} with \eqref{ss5} and substituting $\hat{q}_i$ from \eqref{q_ft}
\begin{align}
%\label{ss6}
%\mathrm{Kn} k^2(\hat{v}_i+\alpha_0 \hat{q}_i)&=f_k\left(\delta_{ik}-\frac{k_ik_k}{k^2}\right)+\mathbbm{i}k_i \mathrm{Kn}(h+\alpha_0 g)
%\\
%\implies
\label{v_ft}
\hat{v}_i=& \medspace \frac{f_j}{\mathrm{Kn}}\left(\frac{\delta_{ij}}{k^2}-\frac{k_i k_j}{k^4}\right)
\nonumber\\
&+\frac{c_p \mathrm{Kn}}{\mathrm{Pr}} \alpha_0^2 f_j\left(\delta_{ij}-\frac{k_i k_j}{k^2}\right)+\mathbbm{i} \frac{k_i h}{k^2}.
\end{align}
Finally,  using Eqs.~\eqref{q_ft} and \eqref{v_ft} in Eq.~\eqref{ccr1_ft}, we obtain \begin{align}
\label{sigma_ft}
\hat{\sigma}_{ij}=&\medspace \mathbbm{i} \, f_\ell 
\left(\frac{k_j \delta_{i\ell}+k_i\delta_{j\ell}}{k^2}-2\frac{k_ik_jk_\ell}{k^4}\right)\nonumber\\
&-2\mathrm{Kn}\left(\frac{k_ik_j}{k^2}-\frac{\delta_{ij}}{3}\right)(h+\alpha_0 g).
\end{align}%
Applying the inverse Fourier transformation in Eqs.~\eqref{theta_ft},  \eqref{p_ft} and \eqref{q_ft}--(\ref{sigma_ft}) with the help of the formulas derived in Appendix\,\ref{app:A},  the field variables turn out to be
\begin{align}
\label{v}
v_i=&\frac{f_j}{\mathrm{Kn}}\left(\frac{x_i x_j}{4\pi r^2}-\frac{2\ln{r}-1}{8\pi}\delta_{ij}\right)\nonumber
\\&+\frac{c_p \mathrm{Kn}}{\mathrm{Pr}} \alpha_0^2 \frac{f_j}{2\pi} \left(\frac{2x_i x_j}{r^4}-\frac{\delta_{ij}}{r^2}\right)+\frac{h x_i}{2\pi r^2},
\\[1ex]
\label{q}
q_i=&\frac{g}{2\pi} \frac{x_i}{r^2}-\frac{c_p \mathrm{Kn}}{\mathrm{Pr}} \alpha_0 \frac{f_j}{2\pi}\left(\frac{2x_i x_j}{r^4}-\frac{\delta_{ij}}{r^2}\right),
\\[1ex]
\label{p}
p=&\frac{f_i x_i}{2\pi r^2},
\\[1ex]
\label{theta}
T=&-\frac{\mathrm{Pr}}{c_p\mathrm{Kn}} \frac{g\,\ln{r}}{2\pi},
\\[1ex]
\label{sigma}
\sigma_{ij}=&\frac{f_\ell x_\ell+2\mathrm{Kn}(h+\alpha_0 g)}{2\pi}\left(\frac{2x_ix_j}{r^4}-\frac{\delta_{ij}}{r^2}\right),
\end{align}
where $r=|x_i|$ and $i,j,\ell \in \{1,2\}$.
The field variables in Eqs.~\eqref{v}--\eqref{sigma} are the fundamental solutions of the linearized CCR model in 2D.
These fundamental solutions in the vectorial/tensorial notation are written as
\begin{align}
\label{vel}
\bm{v}(\bm{r})&=\frac{\bm{f}\cdot\bm{A}(\bm{r})}{8\pi \mathrm{Kn}}+\frac{1}{2\pi}\frac{c_p \mathrm{Kn}}{\mathrm{Pr}} \alpha_0^2  \bm{f}\cdot \bm{B}(\bm{r})
+\frac{h\,\bm{r}}{2\pi r^2},
%+\frac{c_p \mathrm{Kn}}{2\pi \mathrm{Pr}}\alpha_0^2 \bm{f}\cdot \bm{B}(\bm{r}),
\\
\label{pressure}
p(\bm{r})&=\frac{\bm{f}\cdot\bm{r}}{2\pi r^2},
\\
\label{stress}
\bm{\sigma}(\bm{r})&=\frac{\bm{f}\cdot\bm{r}+2 \mathrm{Kn}(h+ g\alpha_0)}{2\pi} \bm{B}(\bm{r}),
\\
\label{temp}
T(\bm{r})&=-\frac{\mathrm{Pr}}{c_p\mathrm{Kn}} \frac{g\,\ln{r}}{2\pi},
\\
\label{heat}
\bm{q}(\bm{r})&=\frac{g}{2\pi}\frac{\bm{r}}{{r}^2}
- \frac{1}{2\pi}\frac{c_p \mathrm{Kn}}{\mathrm{Pr}} \alpha_0  \bm{f}\cdot \bm{B}(\bm{r}),
%-\frac{c_p \mathrm{Kn}}{2\pi \mathrm{Pr}}\alpha_0 \bm{f}\cdot \bm{B}(\bm{r}),
\end{align}
where $r=|\bm{r}|$ and
\begin{align}
\bm{A}(\bm{r})&=\frac{2\bm{rr}}{r^2}-(2\ln{r}-1)\bm{I},
\\ 
\label{abbr}
\bm{B}(\bm{r})&=\frac{2\bm{rr}}{r^4}-\frac{\bm{I}}{r^2}.
\end{align}
Note that, in Eqs.~\eqref{vel}--\eqref{abbr},
\begin{align}
\label{indicial_to_vec}
\left.
\begin{gathered}
\bm{f} = \begin{bmatrix}
f_1 \\ f_2
\end{bmatrix},
\quad
\bm{v}(\bm{r}) = \begin{bmatrix}
v_1(\bm{r})
\\
v_2(\bm{r})
\end{bmatrix},
\quad
\bm{q}(\bm{r}) = \begin{bmatrix}
q_1(\bm{r})
\\
q_2(\bm{r})
\end{bmatrix},
\\[2ex]
\bm{\sigma}(\bm{r}) = \begin{bmatrix*}[r]
\sigma_{11}(\bm{r}) & \sigma_{12}(\bm{r})
\\
\sigma_{12}(\bm{r}) & -\sigma_{11}(\bm{r})
\end{bmatrix*},
\quad
\bm{I} = \begin{bmatrix}
1 & 0
\\
0 & 1
\end{bmatrix}.
\end{gathered}
\right\}
\!
\end{align}
%\begin{align}
%\label{indicial_to_vec}
%\left.
%\begin{aligned}
%\bm{v}(\bm{r}) &= \begin{bmatrix}
%v_1(\bm{r})
%\\
%v_2(\bm{r})
%\end{bmatrix},&
%%\quad
%\bm{\sigma}(\bm{r}) &= \begin{bmatrix*}[r]
%\sigma_{11}(\bm{r}) & \sigma_{12}(\bm{r})
%\\
%\sigma_{12}(\bm{r}) & -\sigma_{11}(\bm{r})
%\end{bmatrix*},
%\\[2ex]
%\bm{q}(\bm{r}) &= \begin{bmatrix}
%q_1(\bm{r})
%\\
%q_2(\bm{r})
%\end{bmatrix},&
%\bm{I} &= \begin{bmatrix}
%1 & 0
%\\
%0 & 1
%\end{bmatrix}.
%\end{aligned}
%\right\}
%\end{align}
%
It is also worthwhile noticing that the fundamental solutions for the linearized NSF and G13 equations in 2D can be obtained directly from Eqs.~\eqref{vel}--\eqref{heat} by taking $\alpha_0 = 0$ and $\alpha_0 = 2/5$, respectively.
The fundamental solutions \eqref{vel}--\eqref{heat} need to be implemented with appropriate boundary conditions for a given problem. 
We shall discuss their implementation and validation in Sec.\,\ref{sec:prob1} and Sec.\,\ref{sec:prob2} for two different problems.

\section{\label{sec:system}Boundary discretization}
\begin{figure}[b!]
\centering
\includegraphics[width=0.9\columnwidth]{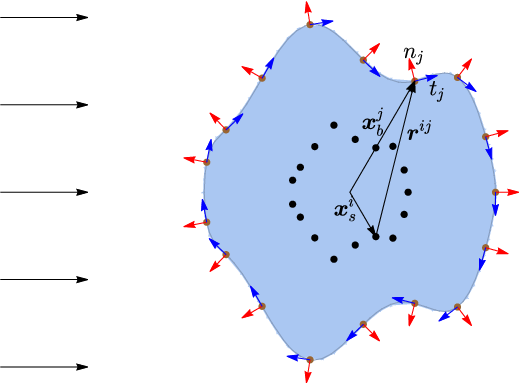}
\caption{\label{fig:arb} Schematic diagram of a flow past an object of an arbitrary shape depicting the 
boundary discretization and the placement of singularities outside the flow domain. 
The red and blue arrows at each boundary node depict the normal (pointing toward the flow domain) and tangential directions at that node, respectively.} 
\end{figure}

%Before addressing the actual validation problems, 
We describe the construction of a system of algebraic equations for applying the MFS through the problem of flow past a complex geometry as depicted in Fig.~\ref{fig:arb}. 
As an example, the geometry of the object in Fig.~\ref{fig:arb} is mathematically defined in the parametric form as
\begin{align}
\label{parametric}
(x,y)=\left(\frac{5}{4} a \cos{\theta}, \frac{1}{4}a (5 - \cos{5\theta}) \sin{\theta}\right)
%\quad 0\leq \theta\leq 2\pi
\end{align}
with $0\leq \theta\leq 2\pi$ and $a\leq 1$ being the dilation factor.
In the MFS, it is quite natural to place the singularity points outside the flow domain \cite{KA1964}. 
%To study the flow outside this geometry, we need to consider the singularity points inside the geometry. 
Notwithstanding, 
the location of the singularity points is a major concern as the results obtained from the MFS are highly sensitive toward the location of singularities \cite{Alves2009, CKL2016, CH2020}. 
% As evident from the results in literature \citep{CH2020,LCK2013}, 
There are two most common ways of distributing singularities in the MFS. 
One way is to place the singularities on a fictitious boundary of a very simple shape---irrespective of the shape of the object---with just one parameter to control; for example, on a circle in the two-dimensional case and on a sphere in the three-dimensional case, and the radius of the circle or sphere would be the controlling parameter. 
Another way is to 
%place the singularities  
recreate a dilated (or shrunk) fictitious boundary, which has the same shape as the boundary of the original object and to place the singularities on this fictitious boundary \cite{CH2020, LCK2013}---similarly to that shown in Fig.~\ref{fig:arb} as well. 
The latter is also easy if the original boundary of the object can be described by a set of parametric equations having only a single controlling parameter, the dilation factor.
% Alternatively, a set of dispersed nodes with no pattern can be used. The goal of this technique is for the user to be able to alter the node density around areas where severe gradients are expected. More details related to optimal source locations can be found in \citep{CH2020,LCK2013}.
For the problem depicted in Fig.~\ref{fig:arb}, we have taken the fictitious boundary to be of the same shape as the original boundary.
%singularities can either be placed as same shape of original boundary, or in case of very complex geometry, singularities can be placed at a circular fictitious boundary outside the domain.

Let $B$ be the number of the discretized boundary nodes and $S$ the number of singularity points. The boundary nodes and the singularities are placed at equispaced angles $\theta$ on the original and the fictitious boundary, respectively, and the distance between both boundaries can be varied by changing the value of the dilation factor $a$.
It may be noted that singularities need not be placed at equispaced angles in principle; nonetheless, we have done so for the sake of simplicity. 
Let $\bm{x}_i^s$ and $\bm{x}_j^b$  be the position vectors of the $i^\mathrm{th}$ singularity site and the $j^\mathrm{th}$ boundary node, respectively, 
then the position vector from the $i^\mathrm{th}$ singularity site to any position $\bm{x}$ in the domain is $\bm{r}_i=\bm{x}-\bm{x}_i^s$ and the position vector from the $i^\mathrm{th}$ singularity site to the $j^\mathrm{th}$ boundary node is $ \bm{r}_{ij}=\bm{x}_j^b-\bm{x}_i^s$.
It is important to note that the subscripts `$i$' and `$j$' are now being used for denoting the $i^\mathrm{th}$ singularity site and $j^\mathrm{th}$ boundary node and consequently, the repetition of indices henceforth shall \emph{not} imply the Einstein summation per se, unless stated otherwise (particularly, in Appendix~\ref{app:A}, wherein the Einstein summation does hold over the repeated indices).
Since the point sources $\bm{f}$, $g$ and $h$---of different strengths---are to be put at each singularity site, there are four degrees of freedom corresponding to each singularity point [two scalars $g$ and $h$ from the point heat and mass sources, and two components $f_1$ and $f_2$ of the point force vector $\bm{f}=(f_1,f_2)^\mathsf{T}$]. 
In total, we have $4\times S$ unknowns, which are determined typically by satisfying the boundary conditions at the boundary points. 
Once the location of the singularity points is decided, the next step in the implementation of the MFS is superposition of the fundamental solutions associated with each singularity site, which makes sense because of the linearity of equations and gives the value of the field variables at the $j^\mathrm{th}$ boundary node. 
%Superposition of the fundamental solutions associated with each singularity site gives the value of the field variables at the $j^\mathrm{th}$ boundary node. 
Superimposing the fundamental solutions \eqref{vel}--\eqref{heat} for each singularity site, the field variables at the $j^\mathrm{th}$ boundary node read
\begingroup
\allowdisplaybreaks
\begin{align}
\label{vj}
\bm{v}_j%(\bm{r}_{ij})
&=\sum_{i=1}^{S}\bigg[\frac{\bm{f}_i\cdot\bm{A}(\bm{r}_{ij})}{8\pi \mathrm{Kn}} + \frac{1}{2\pi}\frac{c_p \mathrm{Kn}}{\mathrm{Pr}}\alpha_0^2 \bm{f}_i\cdot \bm{B}(\bm{r}_{ij}) + \frac{h_i \,\bm{r}_{ij}}{2\pi r_{ij}^2}\bigg],
\\
\label{Pj}
p_j%(\bm{r}_{ij})
&=\sum_{i=1}^{S}\frac{\bm{f}_i\cdot \bm{r}_{ij}}{2\pi r_{ij}^2},
\\
\bm{\sigma}_j%(\bm{r}_{ij})
&=\sum_{i=1}^{S}\frac{\bm{f}_i\cdot \bm{r}_{ij} + 2 \mathrm{Kn}\,(h_i+g_i\,\alpha_0)}{2\pi} \bm{B}(\bm{r}_{ij}),
\\
\label{Tj}
T_j%(\bm{r}_{ij})
&=-\sum_{i=1}^{S}\frac{\mathrm{Pr}}{c_p \mathrm{Kn}} \frac{g_i \, \ln{r}_{ij}}{2\pi},
\\
\label{qj}
\bm{q}_j%(\bm{r}_{ij}) 
&=\sum_{i=1}^{S} \bigg[\frac{g_i}{2\pi}\frac{\bm{r}_{ij}}{r_{ij}^2} 
- \frac{1}{2\pi} \frac{c_p \mathrm{Kn}}{\mathrm{Pr}}\alpha_0 \bm{f}_i\cdot \bm{B}(\bm{r}_{ij}) \bigg],
%\end{aligned}
%\right\}
\end{align}
\endgroup
where $r_{ij}=|\bm{r}_{ij}|$; $\bm{f}_i=({f_1}_i,{f_2}_i)^\mathsf{T}$, $g_i$ and $h_i$ are the point force (vector), point heat source and point mass source, respectively, applied on the $i^\mathrm{th}$ singularity site; and
\begin{align}
%\bm{f}_i=&({f_1}_i,{f_2}_i)^\mathsf{T},\quad r_{ij}=|\bm{r}_{ij}|, 
%\\
\bm{A}(\bm{r}_{ij})=&\frac{2\bm{r}_{ij}\bm{r}_{ij}}{r_{ij}^2}-(2\ln{r}_{ij}-1)\bm{I},
\\ 
\bm{B}(\bm{r}_{ij})=&\frac{2\bm{r}_{ij}\bm{r}_{ij}}{r_{ij}^4}-\frac{\bm{I}}{r_{ij}^2}.
\end{align}
This system is solved for the unknowns ${f_1}_i,{f_2}_i,g_i,h_i$, $i\in\{1,2,3,\dots,S\}$ by employing the boundary conditions at each boundary node. 
Once the unknowns ${f_1}_i,{f_2}_i,g_i,h_i$ for $i\in\{1,2,3,\dots,S\}$ are found, 
the flow variables at any position $\bm{x}$ in the flow domain can be determined 
%by means of superposition of solutions obtained 
simply by dropping the subscript `$j$' everywhere %and replacing $\bm{r}_{ij}$ with $\bm{r}_i$ 
in Eqs.~\eqref{vj}--\eqref{qj}. 
For instance, the velocity $\bm{v}\equiv\bm{v}(\bm{x})$ at a position $\bm{x}$ in the flow domain is given by
\begin{align}
\bm{v}&=\sum_{i=1}^{S}\bigg[\frac{\bm{f}_i\cdot\bm{A}(\bm{r}_{i})}{8\pi \mathrm{Kn}}+\frac{h_i}{2\pi}\frac{\bm{r}_{i}}{r_{i}^2}+\frac{c_p \mathrm{Kn}\alpha_0^2}{2\pi \mathrm{Pr}} \bm{f}_i\cdot \bm{B}(\bm{r}_{i})\bigg].\!
\end{align}
The other flow variables are obtained from Eqs.~\eqref{Pj}--\eqref{qj} analogously. 
%{\color{red}it is straightforward to compute the field variables at any position in the flow domain by means of Eqs.~\eqref{vel}--\eqref{heat}.}
The above procedure to evaluate flow variables %at the $j^\mathrm{th}$ node 
works for any geometry and we have implemented this in a numerical framework. 
We shall elaborate on the placement of boundary nodes and source points, formation and solution of the system separately corresponding to the two problems in Sec.\,\ref{sec:prob1} and Sec.\,\ref{sec:prob2}.
%which will be shown in further application cases.
\section{\label{sec:prob1}Vapor flow confined between two coaxial cylinders}
For the validation of the developed numerical framework, we revisit the problem of a rarefied vapor flow confined between two concentric cylinders. 
The same problem was investigated by Onishi \cite{Onishi1977} with the linearized BGK model and the diffuse reflection boundary conditions.
\subsection{\label{prob1_disc}Problem description}
\begin{figure}[!t]
\centering
\includegraphics[width=\columnwidth]{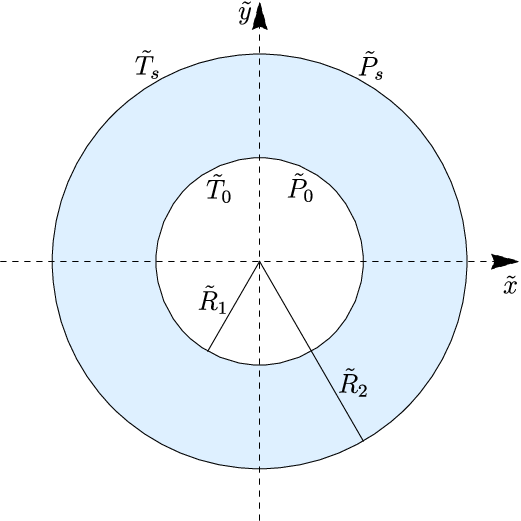} 
\caption{\label{fig:coaxial_schematic}Cross-sectional view of the vapor flow confined between two coaxial cylinders.} 
\end{figure}

Let us consider a moderately rarefied vapor confined between the condensed phases of two concentric infinitely long circular cylinders of radii $\tilde{R}_1$ and $\tilde{R}_2$, where $\tilde{R}_1<\tilde{R}_2$. Owing to the axial symmetry along $\tilde{z}$ axis, it is sufficient to investigate the problem in 2D.
A cross-sectional (two-dimensional) view of the problem is illustrated in Fig.~\ref{fig:coaxial_schematic}.
For the purpose of non-dimensionalization, we take the inner radius as the characteristic length $\tilde{L}$, i.e.~$\tilde{L}=\tilde{R}_1$.
Consequently, the dimensionless radii of the inner and outer cylinders are $r_1=\tilde{R}_1/\tilde{L}=1$ and $r_2=\tilde{R}_2/\tilde{L}$, respectively.
The condensed phases of the vapor at the inner and outer cylinders are assumed to be negligibly thin.
Let the temperatures of the inner and outer condensed phases be maintained at uniform temperatures $\tilde{T}_0$ and $\tilde{T}_s$, respectively. 
Moreover, let the saturation pressures of the condensed phases corresponding to the temperatures $\tilde{T}_0$ and $\tilde{T}_s$ be $\tilde{P}_0$ and $\tilde{P}_s$, respectively; see Fig.~\ref{fig:coaxial_schematic}.
Again, for the purpose of linearization and non-dimensionalization, we take the temperature at the inner wall $\tilde{T}_0$  as the reference temperature and the saturation pressure at the inner wall $\tilde{P}_0$ as the reference pressure.
Thus, the dimensionless perturbations in the temperature and saturation pressure at the inner wall vanish, and the dimensionless perturbations in the temperature and saturation pressure at the outer wall read
\begin{align}
\tau_s=\frac{\tilde{T}_s-\tilde{T}_0}{\tilde{T}_0}
\quad\text{and}\quad
p_s=\frac{\tilde{P}_s-\tilde{P}_0}{\tilde{P}_0},
\end{align}
respectively.
\subsection{Analytic solution of Onishi \cite{Onishi1977}}
Onishi \cite{Onishi1977} investigated the problem by employing an asymptotic theory \cite{Sone2002}.
%We assume two indefinitely long coaxial cylinders separated by a pure vapor gas. Owing to symmetry, it is sufficient to study the flow features in a two-dimensional cross-section.
 %So for simplification, we investigate a two-dimensional cross-section of this problem with condensed phase boundary conditions.
According to this theory,
a field variable $\tilde{h}$ of the gas can be written as
\begin{align}
\tilde{h}=\tilde{h}_H+\tilde{h}_K,
\end{align}
where $\tilde{h}_H$ is referred to as the hydrodynamic part or the Hilbert part that describes the flow behavior in the bulk of the domain and $\tilde{h}_K$ is referred to as the kinetic boundary layer part or the Knudsen layer part that can be seen as a correction to the Hilbert part and is significant only in a small layer near an interface. 
Both $\tilde{h}_H$ and $\tilde{h}_K$ for all field variables are expanded in power series of the Knudsen number, and the contribution at each power of the Knudsen number is then computed by means of the considered model (the BGK model in [\onlinecite{Onishi1977}]) and appropriate boundary conditions (the diffuse reflection boundary conditions in [\onlinecite{Onishi1977}]).

%The Knudsen layer is a rarefaction phenomenon that occurs when a solid surface is separated by a few mean free paths. Molecule-surface collisions are more prominent in this stratum than intermolecular collisions. This causes velocity slip and the development of non-Newtonian behavior, in which the Knudsen layer begins to dominate the flow behavior. 
%The terms are expanded in power series of Knudsen number. 
The linearized CCR model is anyway not able to predict Knudsen layers. Therefore, it makes sense to compare the results obtained from the MFS only with the Hilbert part of the solution given in Ref.~[\onlinecite{Onishi1977}].
For the problem under consideration and for the linearized BGK model with the diffuse reflection boundary conditions, the Hilbert part of the solution is indeed straightforward to determine by solving a set of simple ordinary differential equations analytically, see Ref.~[\onlinecite{Onishi1977}].
Denoting the radius ratio by $\beta=r_2/r_1$ and the ratio of $p_s$ to $\tau_s$ by
%the slope of the saturated vapor pressure-temperature curve at temperature $\tilde{T}_0$ by 
$\gamma=p_s/\tau_s$, the analytic solution obtained from the linearized BGK model with the diffuse reflection boundary conditions for $\mathrm{Kn}\approx 0$ is given by \cite{Onishi1977}
\begin{align}
\label{analy_p}
p=& \medspace
p_s\left(\frac{1}{r_1}+\frac{1}{r_2}\right)^{-1}\frac{1}{r_1},
\\
\label{analy_vr}
v_r=&-\frac{p_s}{C_0}\left(\frac{1}{r_1}+\frac{1}{r_2}\right)^{-1}\frac{1}{r},
\\
\label{analy_T}
T=& \medspace
\tau_s\left[\left(1-\frac{D_0}{C_0}\gamma\right)\frac{\ln{r}}{\ln{\beta}}-\left(1-\frac{D_0}{C_0}\gamma\right)\frac{\ln{r}_1}{\ln{\beta}}\right]
\nonumber\\
&+\frac{D_0}{C_0}\gamma\tau_s\left(\frac{1}{r_1}+\frac{1}{r_2}\right)^{-1}\frac{1}{r_1},
\\
\label{analy_q}
q_r=&0,
\end{align}
where $C_0=2.132039$ and $D_0=0.4467494$.

\subsection{\label{prob1_bcs}Boundary conditions and implementation of the MFS}
We shall revisit the problem described above by means of the MFS applied on the linearized CCR model. 
Recall that we have already determined the fundamental solutions of the linearized CCR model and outlined the way to implement them in Sec.\,\ref{sec:solutions} for a general two-dimensional object.
The solution for the field variables at the $j^\mathrm{th}$ boundary node  (Eqs.~\eqref{vel}--\eqref{heat}) can directly be used once the boundary nodes and singularity points for the present problem have been decided.

\begin{figure}[!t]
\centering
\includegraphics[width=\columnwidth]{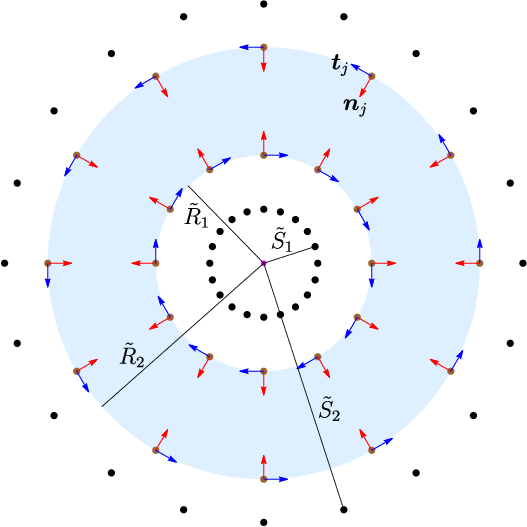}
\caption{\label{fig:coaxial_discr}Schematic of the boundary nodes on the boundaries and singularity points outside the flow domain for the problem illustrated in Fig.~\ref{fig:coaxial_schematic}. The red and blue arrows at each boundary node depict the normal (pointing toward the flow domain) and tangential directions at that node, respectively.%
} 
\end{figure}
Since the singularity sites are to be placed outside of the computational domain, we assume the source points to be placed on two fictitious circular boundaries, one inside the circle associated with the inner cylinder and the other outside the circle associated with the outer cylinder, as shown in Fig.~\ref{fig:coaxial_discr}. 
Note that both fictitious boundaries are concentric with the circles associated with the cylinders.
Let the radii of the inner and outer fictitious boundaries be $\tilde{S}_1$ and $\tilde{S}_2$, respectively. %
For simplicity, we consider $N_s$ equispaced source points on each of the two fictitious boundaries and $N_b$ equispaced boundary nodes on each of the actual boundaries (the boundaries of the inner and outer cylinders).
As explained in Sec.\,\ref{sec:system}, we have four degrees of freedom corresponding to each source point, and the total number of singularity points for the problem under consideration is $S=2N_s$. Thus, there will be a total of $4\times S = 4\times2N_s$ unknowns in the problem.
Accordingly, the summations in Eqs.~\eqref{vel}--\eqref{heat} will run from $i=1$ to $i=2N_s$.

Boundary conditions at the $j^{\mathrm{th}}$ boundary node are obtained from \eqref{bc_1}--\eqref{bc_3} by replacing the flow variables and the normal and tangent vectors with their respective values at the $j^{\mathrm{th}}$ boundary node.
%(already calculated in Sec.\,\ref{sec:system}).
Furthermore, since the walls of the cylinders are fixed, $\bm{v}^I=\bm{0}$. 
Consequently, the boundary conditions at the $j^{\mathrm{th}}$ boundary node read
\begin{align}
\label{bc_11}
\bm{v}_j\cdot \bm{n}_j=& -\eta_{11}(p_j-p_{\mathrm{sat}} + \bm{n}_j\cdot\bm{\sigma}_j\cdot\bm{n}_j)\nonumber
\\&+\eta_{12}(T_j-T^I+\alpha_0 \bm{n}_j\cdot\bm{\sigma}_j\cdot\bm{n}_j),
\\[2ex]
\label{bc_21}
\bm{q}_j\cdot \bm{n}_j=& \medspace
\eta_{12}(p_j-p_{\mathrm{sat}} + \bm{n}_j\cdot\bm{\sigma}_j\cdot\bm{n}_j)\nonumber\\&-(\eta_{22}+2\tau_0)(T_j-T^I+\alpha_0 \bm{n}_j\cdot\bm{\sigma}_j\cdot\bm{n}_j),
\end{align}
\begin{align}
\label{bc_31}
\bm{t}_j\cdot\bm{\sigma}_j\cdot\bm{n}_j=&-\varsigma (\bm{v}_j+\alpha_0 \bm{q}_j)\cdot\bm{t}_j.
\end{align}
%As described in \ref{prob1},
The dimensionless perturbations in saturation pressures at the inner and outer interfaces are $p_{\mathrm{sat}}=0$ and $p_{\mathrm{sat}}=p_s$, respectively, and the dimensionless perturbations in temperatures at the inner and outer interfaces are $T^I=0$ and $T^I=\tau_s$, respectively, which need to be replaced in boundary conditions \eqref{bc_11}--\eqref{bc_31} accordingly.
%The saturation pressure $p_{\mathrm{sat}}$ at the inner and outer cylinders are $0$ and $p_s$ respectively. The temperature at the inner and outer interfaces $T^I$ are $0$ and $\tau_s$, respectively. 
Note that boundary conditions \eqref{bc_11}--\eqref{bc_31} are to be satisfied at $B=2N_b$ boundary nodes. 
On substituting the values of the field variables at the $j^\mathrm{th}$ boundary node from \eqref{vj}--\eqref{qj} into boundary conditions \eqref{bc_11}--\eqref{bc_31}, the resulting system of equations (associated with the $j^\mathrm{th}$ boundary node) can be written in a matrix form as
\begin{align}
\label{matrix}
\sum_{i=1}^{S} M_{ji} \bm{u}_i=\bm{b}_j,
\end{align}
for the unknown vector associated with the $i^\mathrm{th}$ singularity $\bm{u}_i=({f_1}_i,{f_2}_i,g_i,h_i)^\mathsf{T}$.
Here, $M_{ji}$'s are coefficient matrices of dimensions  $3\times 4$ and $\bm{b}_j$ is the $3\times 1$ vector containing the interface properties, such as $p_s$ and $\tau_s$. 
We collect all such $B$ systems into a new system
\begin{align}
\label{collocation_system}
\mathcal{M} \bm{\mathcal{X}}=\bm{\mathcal{B}},
\end{align}
where $\bm{\mathcal{X}} = \big({f_1}_1,{f_2}_1,g_1,h_1,{f_1}_2,{f_2}_2,g
_2,h_2,\dots,{f_1}_S,{f_2}_S,\allowbreak g_S,h_S\big)^\mathsf{T}$ is the vector containing all $4S$ unknowns, the matrix $\mathcal{M}$---containing the coefficients of the unknowns---has dimensions $3B\times 4S$ (or $6N_b\times 8N_s$) and is referred to as the collocation matrix.
We have solved system \eqref{collocation_system} in the computer algebra software, Mathematica\textsuperscript{\textregistered} using the method of least squares. 
For the identification purpose, the first $N_s$ singularity points ($i=1,2,\dots,N_s$) in our code belong to the inner fictitious boundary and the rest $N_s$ singularity points ($i=N_s+1,N_s+2,\dots,2N_s$) to the outer fictitious boundary.
Similarly, the first $N_b$ boundary nodes ($j=1,2,\dots,N_b$) belong to the actual inner boundary and the rest $N_b$ boundary nodes ($j=N_b+1,N_b+2,\dots,2N_b$) to the actual outer boundary.
\subsection{Results and discussion}
%We have implemented the MFS for the given problem using the lenearized CCR model and validated our results with those obtained with the linearized BGK model in Ref.~[\onlinecite{Onishi1977}]. 
For numerical computations, we have taken $N_b=100$ boundary nodes on each of the actual boundaries and $N_s=100$ singularity points on each of the fictitious boundaries. 
%The results for the MFS have been obtained by taking the number of boundary nodes $N_b=50$ on each of the original boundaries and number of singularity points $N_s=50$ on each of the fictitious boundaries. 
The dimensionless radii of the original and fictitious boundaries are taken as $r_1=1$, $r_2=2$, $s_1=\tilde{S}_1/\tilde{R}_1=0.5$ and $s_2=\tilde{S}_2/\tilde{R}_1=4$. 
%The Prandtl number for the BGK model is unity %$\mathrm{Pr}=1$  \cite{Struchtrup2005}.
% and the parameter $\alpha_0$ is taken as $0.3197$, the value of $\alpha_0$ for hard spheres \cite{RGS2018}. 
%The parameters $\mathrm{Pr}$ and $\alpha_0$ have been taken as $1$ and $0.31$, respectively. 

\begin{figure}[!t]
\centering
\includegraphics[width=\columnwidth]{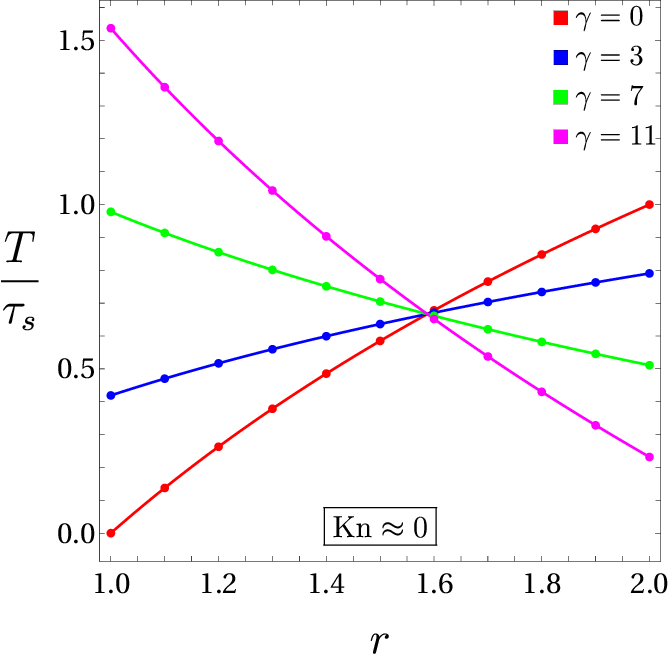} 
\caption{\label{fig:temp_gamma}
Variation of the (scaled) temperature in the gap between the two cylinders for different values of $\gamma$. 
The solid lines denote the results obtained from the MFS applied on the CCR model and the symbols indicate the analytic results from Eq.~\eqref{analy_T}, which was obtained analytically from the linearized BGK model for $\mathrm{Kn} \approx 0$ in Ref.~[\onlinecite{Onishi1977}].
The other parameters are $N_b=100$, $N_s=100$, $r_1=1$, $r_2=2$, $s_1=0.5$ and $s_2=4$.
}
\end{figure}

Figure~\ref{fig:temp_gamma} illustrates the variation of the (scaled) temperature of the vapor in the radial direction for $\mathrm{Kn} \approx 0$ and for different values of the parameter $\gamma \,(=p_s/\tau_s)$, wherein $\tau_s=4$ is fixed and $p_s$ is being varied for varying $\gamma$. 
The solid lines represent the results obtained from our numerical framework based on the MFS while the symbols delineate the results from Eq.~\eqref{analy_T}, which was obtained analytically for $\mathrm{Kn} \approx 0$ through an asymptotic theory \cite{Sone2002} performed on the linearized BGK model in Ref.~[\onlinecite{Onishi1977}]. 
It is evident from the figure that the results obtained with the MFS in the present work are in an excellent agreement with the analytic results  from the linearized BGK model for $\mathrm{Kn} \approx 0$.
\begin{figure}[!t]
\centering
\includegraphics[width=\columnwidth]{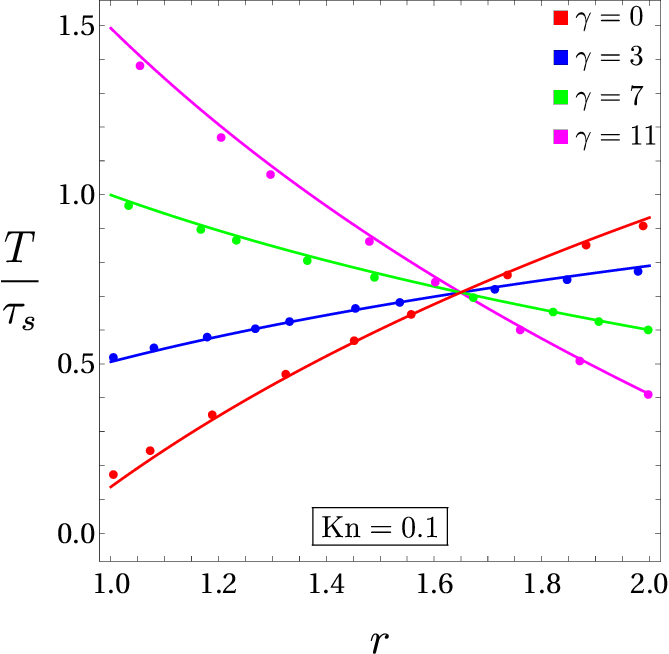} 
\caption{\label{fig:temp_gamma_Kn0p1}Same as Fig.~\ref{fig:temp_gamma} except for the symbols denote the data from Ref.~[\onlinecite{Onishi1977}] for $\mathrm{Kn} = 0.1$ obtained using the linearized BGK model.
}
\end{figure}
%
%\begin{widetext}
%\begin{minipage}{\linewidth}
%\begin{figure}%[!tb]
%\centering
%\includegraphics[scale=0.6]{figures/gamma0} 
%\\
%\includegraphics[scale=0.6]{figures/gamma3} 
%\\%[2ex]
%\includegraphics[scale=0.6]{figures/gamma7} 
%\\
%\includegraphics[scale=0.6]{figures/gamma11} 
%\caption{\label{fig:temp_gamma}Comparison of the results obtained from the MFS (solid lines) and linearized BGK model (symbols) for the temperature variations between the two cylinders for different values of $\gamma$ (on taking $\tau_s=4$, $N_b=50 ,\, N_s=50,\, r_1=1,\, r_2=2,\, s_1=0.5,\, s_2=4 $).}
%\end{figure}
%\end{minipage}
%\end{widetext}
%
%
%\begin{figure}[!h]
%\centering
%\includegraphics[height=60mm]{figures/vel_coaxial}
%\caption{\label{fig:vel_coaxial}Variation of the (a) radial heat flux and (b) radial velocity with $\gamma$.}
%\end{figure}
%%
%\begin{figure}[!h]
%\centering
%\includegraphics[height=60mm]{figures/heat_coaxial} \hfill
%\caption{\label{fig:heat_coaxial}Variation of the (a) radial heat flux and (b) radial velocity with $\gamma$.}
%\end{figure}
Although not shown here for brevity, the results for the pressure and velocity from the MFS are also in excellent agreement with the analytic results from Eqs.~\eqref{analy_p} and \eqref{analy_vr} for $\mathrm{Kn} \approx 0$. 
It is also evident from Fig.~\ref{fig:temp_gamma} that the temperature increases on moving away from the inner cylinder toward the outer cylinder for smaller values of $\gamma$ (red and blue curves and symbols in the figure) and vice versa for larger values of $\gamma$ (green and magenta curves and symbols in the figure).
This indicates the existence of a reverse temperature gradient after a  critical value of $\gamma$. 
Indeed, at this critical value of $\gamma$, the (scaled) temperature remains constant along the radial direction.
An expression for this critical value of $\gamma$ from the asymptotic theory \cite{Sone2002} is given by \cite{Onishi1977}
\begin{align}
\label{gammac}
\gamma_c=\frac{C_0}{D_0}\left[1-\mathrm{Kn} \frac{C_0}{D_0}(0.124226)\left(\frac{1}{r_1}-\frac{1}{r_2}\right) + \mathcal{O}(\mathrm{Kn}^2)\right].
\end{align}
For $\mathrm{Kn} \approx 0$, the critical value of $\gamma$ from the above expression is $\gamma_c = C_0/D_0 \approx 4.772337$.
From the MFS presented here, the critical value of $\gamma$ for $\mathrm{Kn} \approx 0$ turns out to be $\gamma_c \approx 4.7723$, which is also very close to that computed from the above expression. 
The phenomenon of reverse temperature gradient can be understood form boundary condition \eqref{bc_21} as follows.
There are two factors determining the normal heat flux component in boundary condition \eqref{bc_21} according to which the evaporation/condensation rate depends on (i) the difference between the pressure and saturation pressure, and (ii) the temperature difference between the temperatures of the gas (or vapor) and and interface.
The temperature gradient gets reversed when one dominates the other.

To examine the capabilities of the developed method, we also study the problem for higher Knudsen numbers. 
Figure~\ref{fig:temp_gamma_Kn0p1} exhibits the variation of the (scaled) temperature of the vapor in the radial direction for $\mathrm{Kn} = 0.1$ and for different values of the parameter $\gamma$.
The  solid lines again represent the results obtained from our numerical framework based on the MFS but the symbols now denote the data from the linearized BGK model taken directly from Ref.~[\onlinecite{Onishi1977}].
It is clear from the figure that the results from the MFS are in good agreement with those from the  linearized BGK model even for $\mathrm{Kn} = 0.1$; nonetheless, the quantitative differences in the results from both methods are now noticeable.
In addition, Fig.~\ref{fig:temp_gamma_Kn0p1} also shows the existence of a reverse temperature gradient.
For $\mathrm{Kn} = 0.1$, the critical value of $\gamma$, at which the phenomenon of reverse temperature gradient occurs, computed from the MFS 
%in the case of $\mathrm{Kn} = 0.1$ 
is $\gamma_c=4.66247$ whereas its reported value from the linearized BGK model in Ref.~[\onlinecite{Onishi1977}] is $\gamma_c=4.63087$.

\begin{figure}[!t]
\centering
\includegraphics[width=\columnwidth]{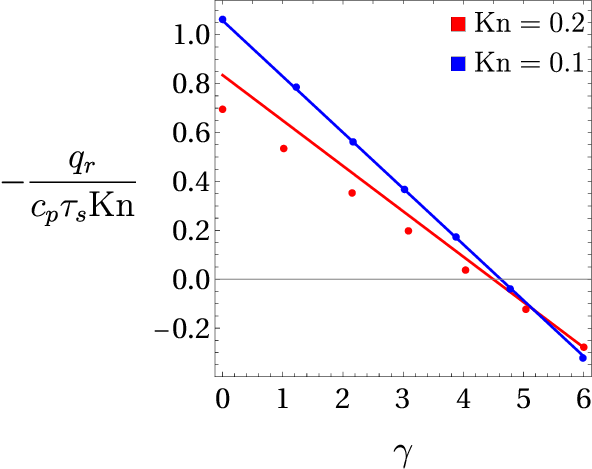}
\caption{
\label{fig:heat_coaxial}%
Variation of the (scaled) radial heat flux with $\gamma$. 
The solid lines denote the results obtained from the MFS applied on the CCR model and the symbols indicate the data taken directly from Ref.~[\onlinecite{Onishi1977}], which were obtained using the linearized BGK model.
%Comparison of the results obtained from the MFS (solid lines) and linearized BGK model (symbols) for the scaled radial heat flux on inner cylinder varying with respect to $\gamma$.
The other parameters are the same as those for Fig.~\ref{fig:temp_gamma}.
} 
\end{figure}
\begin{figure}[!t]
\includegraphics[width=\columnwidth]{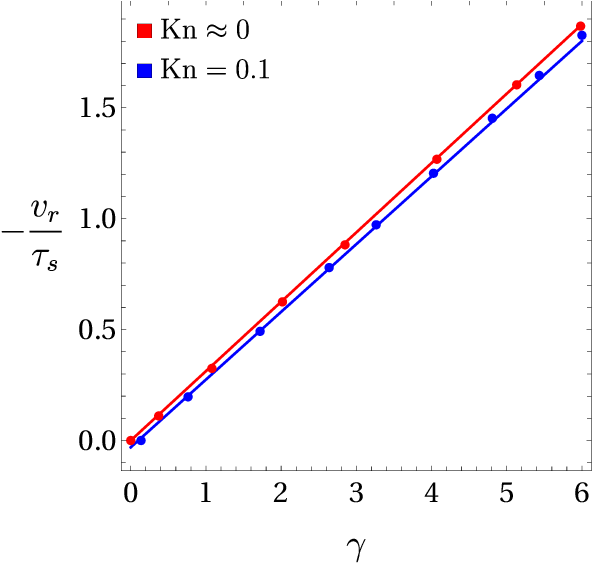}
\caption{\label{fig:vel_coaxial}Variation of the (scaled) radial velocity with $\gamma$.
The solid lines denote the results obtained from the MFS applied on the CCR model and the symbols indicate those from the linearized BGK model (from Eq.~\eqref{analy_vr} in the case of $\mathrm{Kn} \approx 0$ and directly from Ref.~[\onlinecite{Onishi1977}] in the case of  $\mathrm{Kn} = 0.1$).
The other parameters are the same as those for Fig.~\ref{fig:temp_gamma}.}
\end{figure}

To have further insight on the reverse temperature gradient, the (scaled) radial heat flux at the actual inner boundary (i.e.~at $r=1$) is plotted against $\gamma$ in Fig.~\ref{fig:heat_coaxial}.
The solid lines and symbols denote the results from the MFS in the present work and the data from the linearized BGK model given in Ref.~[\onlinecite{Onishi1977}], respectively.
It is apparent from the figure that our results for the radial heat flux are also in good agreement with the data from the linearized BGK model for a smaller value of the Knudsen number ($\mathrm{Kn}=0.1$ in the figure); however, for a higher value of the Knudsen number ($\mathrm{Kn}=0.2$ in the figure), there is a noticeable mismatch between the results obtained from the MFS and the data from the linearized BGK model given in Ref.~[\onlinecite{Onishi1977}]. 
A plausible reason for this discrepancy could be the truncation of power series at the first order  in  Ref.~[\onlinecite{Onishi1977}] because the neglected terms in the series could have significant contributions for larger values of the Knudsen number.
Figure~\ref{fig:heat_coaxial} also shows that for each value of the Knudsen number, there is a $\gamma$ at which the radial heat flux changes its sign. 
This $\gamma$ is indeed the same as the $\gamma_c$ described above, at which reversal of the temperature gradient takes place.
%may be due to negligence of higher order terms in power series of the Knudsen number in  Ref.~[\onlinecite{Onishi1977}] as for higher values of $\mathrm{Kn}$, those terms might have significant effect.

Through the plots of heat flux lines, also not shown here for brevity, it has been found that, in the case of $\tau_s>0$, heat  flows from the outer cylinder toward the inner cylinder for $\gamma<\gamma_c$ and vice versa for $\gamma>\gamma_c$ . 
This makes sense in view of Figs.~\ref{fig:temp_gamma} and \ref{fig:temp_gamma_Kn0p1}.
The direction of heat flow reverses in both cases when $\tau_s$ is taken to be negative or, in other words, when the initial temperature of the inner cylinder is taken higher than that of the outer cylinder.

%Figure~\ref{fig:heat_coaxial} depicts the variation of the scaled dimensionless radial heat flux (scaled in order to compare with Ref.~[\onlinecite{Onishi1977}]) with respect to parameter $\gamma$ (again for fixed $\tau_s=4)$.
%The figure depicts that for both $\mathrm{Kn}=0.1$ and $\mathrm{Kn}=0.2$, heat flux turns out to be negative after the critical value of $\gamma$, due to reverse temperature gradient. 
%The heat flows from cold surface (inner wall) to hot surface (outer wall) for values  $\gamma>\gamma_c$. The direction of heat flow is opposite if $\tau_s$ is taken to be negative or the outer wall is assumed to be cooler than the inner wall. 
%Furthermore, it is found that the results are in good agreement for smaller value of the Knudsen number i.e. $\mathrm{Kn}=0.1$ but for a higher value ($\mathrm{Kn}=0.2$), there is a mismatch between the results obtained from the MFS and the linearized BGK model data. This discrepancy may be due to negligence of higher order terms in power series of the Knudsen number in  Ref.~[\onlinecite{Onishi1977}] as for higher values of $\mathrm{Kn}$, those terms might have significant effect.

Figure~\ref{fig:vel_coaxial} displays the (scaled) radial velocity plotted against $\gamma$ for $\mathrm{Kn} \approx 0$ and $\mathrm{Kn} =0.1$. 
The solid lines are again the results from the MFS in the present work while the symbols in the case of $\mathrm{Kn} \approx 0$ denote the results from Eq.~\eqref{analy_vr} and those in the case of $\mathrm{Kn} = 0.1$ denote the data taken from Ref.~[\onlinecite{Onishi1977}]; nevertheless, in both cases symbols denote the results from the linearized BGK model.  
The figure also demonstrates a good agreement between the results from the method developed in the present work and those from the linearized BGK model.

%The variation of (scaled) radial velocity with respect to $\gamma$ is shown in Fig. \ref{fig:vel_coaxial} and its validation with the linearized BGK data (symbols) clearly demonstrates a good agreement between the two models.

%The negative heat flow phenomenon specified for rarefied gases can be seen in Figure~\ref{fig:heat_vel} a) for values $\gamma>\gamma_c$ as heat flows from cold to hot surface.

\section{\label{sec:prob2}Rarefied gas flow between two non-coaxial cylinders}
In this section, we revisit the problem of flow induced by a temperature difference in a rarefied gas confined between two non-coaxial cylinders via the MFS developed above.
The same problem was investigated numerically by Aoki, Sone and Yano \cite{Aoki_Sone_1989} with the linearized BGK model and the diffuse reflection boundary conditions.
\subsection{Problem description}
%In this problem, flow rendered by a temperature difference in a rarefied gas confined between two non-coaxial cylinders at rest with different uniform temperatures is investigated. The problem has been investigated with the BGK model in [\onlinecite{Aoki_Sone_1989}].
Let us consider a rarefied (monatomic) gas confined between two infinitely long circular cylinders of radii $\tilde{R}_1$ and $\tilde{R}_2$ (with $\tilde{R}_1<\tilde{R}_2$) that are not coaxial.
Again, owing to the axial symmetry, it is sufficient to investigate the problem in 2D. 
\begin{figure}[!t]
\centering
\includegraphics[width=\columnwidth]{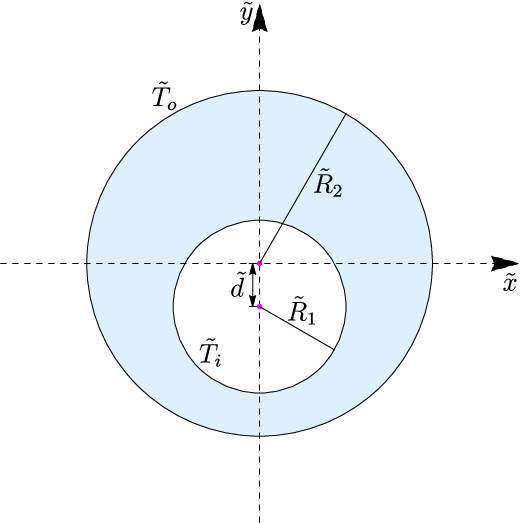} \caption{\label{fig:noncoaxial_schematic}Cross-sectional view of the flow of a rarefied gas confined between two non-coaxial cylinders having different wall temperatures.}
\end{figure}
Let the locations of both cylinders be fixed according to the cross-sectional view portrayed in Fig.~\ref{fig:noncoaxial_schematic} so that the centers of the circles associated with the inner and outer cylinders be at the origin and at $(0, - \tilde{d})$, respectively.  
%the distance between their centers be $\tilde{d}$. 
Furthermore, let the temperatures of the inner and outer cylinders be kept fixed at $\tilde{T}_i = \tilde{T}_0$ and $\tilde{T}_o = \tilde{T}_0 (1+\Delta\tau)$, respectively, with $\Delta\tau$ being sufficiently small in comparison to $\tilde{T}_0$ so that the linear theory remains meaningful.

For the purpose of non-dimensionalization, we again take the radius of the inner cylinder as the characteristic length $\tilde{L}$, i.e.~$\tilde{L}=\tilde{R}_1$.
Consequently, the dimensionless radii of the inner and outer cylinders are $r_1=\tilde{R}_1/\tilde{L}=1$ and $r_2=\tilde{R}_2/\tilde{L}$, respectively, and the dimensionless distance between the centers of the cylinders is $d =\tilde{d}/\tilde{L}$.
Furthermore, for the purpose of the linearization and non-dimensionalization, the equilibrium pressure of the gas $\tilde{p}_0$ is taken as the reference pressure and the temperature of the inner cylinder $\tilde{T}_i$ as the reference temperature so that the dimensionless perturbations in temperatures on the inner and outer walls are $T_i = (\tilde{T}_i - \tilde{T}_i) / \tilde{T}_i=0$ and $T_o = (\tilde{T}_o - \tilde{T}_i) / \tilde{T}_i = \Delta \tau$, respectively.
For comparing the results from the present method with those of Ref.~[\onlinecite{Aoki_Sone_1989}], the parameters are fixed to $r_2 = 2$, $d = 0.5$ and $\Delta \tau = 1$.

\begin{figure}[!t]
\centering
\includegraphics[width=\columnwidth]{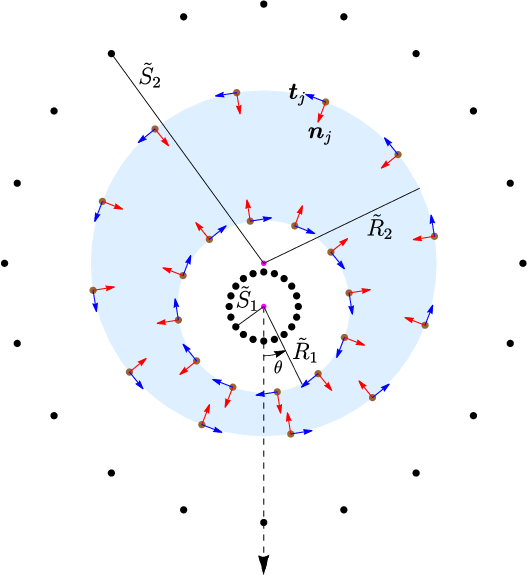} 
\caption{\label{fig:noncoaxial_discr}Schematic of the boundary nodes on the boundaries and singularity points outside the flow domain for the problem illustrated in Fig.~\ref{fig:noncoaxial_schematic}. 
The red and blue arrows at each boundary node depict the normal (pointing toward the flow domain) and tangential directions at that node, respectively.%
}
\end{figure}
\subsection{Boundary Conditions and implementation of the MFS}
In order to place the singularity sites outside the computational domain, we again assume the source points to be placed on two fictitious circular boundaries, one inside the circle associated with the inner cylinder and the other outside the circle associated with the outer cylinder, as shown in Fig.~\ref{fig:noncoaxial_discr}. 
The inner (outer) fictitious boundary is concentric with the circle associated with the inner (outer) cylinder. 
Let the radii of the inner and outer fictitious boundaries be $\tilde{S}_1$ and $\tilde{S}_2$, respectively. 
Consequently, the dimensionless radii of the inner and outer fictitious boundaries are $s_1 = \tilde{S}_1/\tilde{L}$ and $s_2 = \tilde{S}_2/\tilde{L}$.
Similarly to the above, we consider $N_s$ equispaced source points on each of the two fictitious boundaries and $N_b$ equispaced boundary nodes on each of the actual boundaries (the boundaries of the inner and outer cylinders).

Since the walls of the cylinders are fixed for this problem as well, $\bm{v}^I=\bm{0}$. 
Hence, the boundary conditions \eqref{bc_11}--\eqref{bc_31} at the $j^{\mathrm{th}}$ boundary node hold true for the present problem as well.
However, since the present problem does not involve evaporation and condensation, the evaporation/condensation coefficient $\vartheta$ is zero for this problem. 
Consequently, boundary conditions \eqref{bc_11}--\eqref{bc_31} for the problem under consideration further reduce to
%
%
%We apply the diffuse-reflection boundary conditions for which the evaporation/condensation coefficient $\vartheta$ is zero. The reduced boundary conditions at the $j^{\mathrm{th}}$ boundary node read
\begin{align}
\label{bc1}
\bm{v}_j\cdot \bm{n}_j&= 0,
\\
\label{bc2}
\bm{q}_j\cdot \bm{n}_j&= -2\tau_0(T_j-T^I+\alpha_0 \,\bm{n}_j\cdot\bm{\sigma}_j\cdot\bm{n}_j),
\\
\label{bc3}
\bm{t}_j\cdot\bm{\sigma}_j\cdot\bm{n}_j&=-\varsigma (\bm{v}_j +\alpha_0^{\prime} \bm{q}_j)\cdot\bm{t}_j.
\end{align}
Note that the coefficient $\alpha_0$ in boundary condition \eqref{bc3} has been changed to $\alpha_0^{\prime}=1/5$ (see, e.g., Refs.~[\onlinecite{Struchtrup2005, ST2008, TTS2009}]) in order to have a fair comparison with the findings of Ref.~[\onlinecite{Aoki_Sone_1989}].
%{\color{red}
%Here, $\zeta$ is the thermal slip coefficient whose value as per literature \citep{Sharipov2004} lies between $0.2$ and $0.31$ (Refer to Sec.\,\ref{app:coef} for details).}
The interface temperature $T^I$ in boundary condition \eqref{bc2} is $0$ for the inner cylinder and $\Delta \tau$ for the outer cylinder. 
%%
%The source points are placed inside the inner cylinder and outside of the outer cylinder. The fictitious boundary for inner source points is placed at the dimensionless radius $s_1(=\tilde{S}_1/\tilde{L})$ concentric to the inner cylinder at center $(0,-d)$ and the outer fictitious boundary is placed at dimensionless radius $s_2(=\tilde{S}_2/\tilde{L})$ concentric to the outer cylinder centered at the origin. Again for convenience, $N_s$ source points are placed on each fictitious boundary and $N_b$ boundary nodes on each original boundary. 

The construction of the collocation matrix and the formation of system \eqref{collocation_system} for the present problem is exactly similar to that demonstrated in Sec.~\ref{prob1_bcs}. 
We have again solved system  \eqref{collocation_system} for the present problem analogously in the computer algebra software, Mathematica\textsuperscript{\textregistered} using the method of least squares to determine the unknowns ${f_1}_1,{f_2}_1,g_1,h_1,{f_1}_2,{f_2}_2,g
_2,h_2,\dots,{f_1}_S,{f_2}_S,\allowbreak g_S,h_S$.
%
%
%its dimensions are the same as that demonstrated in Sec.~\ref{prob1_bcs}.
% After solving the system, we get the values of unknowns  $({f_1}_i,{f_2}_i,g_i,h_i)$ for every $i^\mathrm{th}$ source point. 
%
\subsection{Results and discussion}
\begin{figure}[!t]
\centering
\includegraphics[width=\columnwidth]{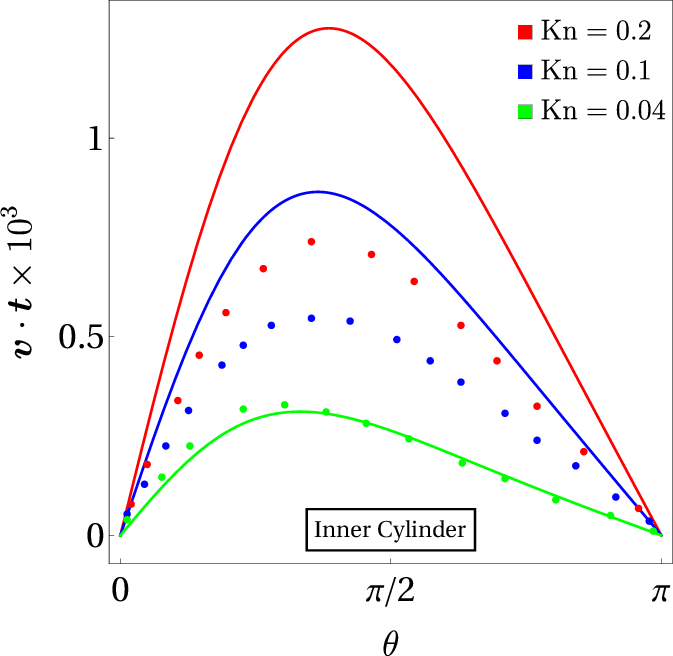} 
\\[3ex]
\includegraphics[width=\columnwidth]{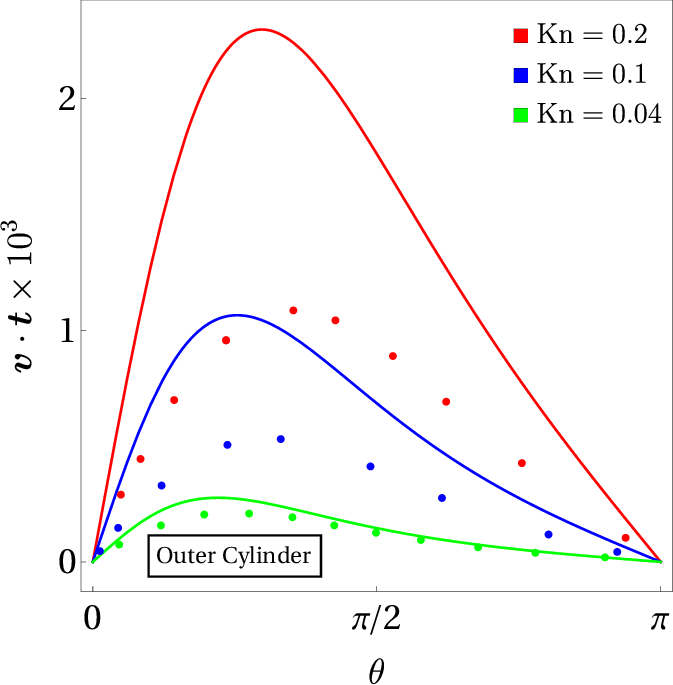}
\caption{\label{fig:vel}%
Tangential velocity on the right halves of the inner and outer circles associated with the respective cylinders plotted against the angle $\theta$ for different values of the Knudsen number and for $\Delta \tau = 1$. 
The solid lines denote the results obtained from the MFS applied on the CCR model and the symbols indicate the data from the linearized BGK model \cite{Aoki_Sone_1989}. 
The other parameters are the same as those for Fig.~\ref{fig:temp_gamma}.
} 
\end{figure}
We have computed the results numerically by taking the parameters as $\Delta\tau=1$, $N_b = N_s = 100$, $r_1=1$, $r_2=2$, $s_1=0.5$ and $s_2=4$. 
In all the figures (except for Fig.~\ref{fig:NS_stream}) below, the solid lines represent the results obtained with the MFS applied on the CCR model in the present work and the symbols denote the data taken from Ref.~[\onlinecite{Aoki_Sone_1989}], which were obtained using the linearized BGK model.

Figure~\ref{fig:vel} illustrates the variation of the tangential component of the (dimensionless) velocity on the right halves of the inner (top row) and outer (bottom row) circles associated with the respective cylinders with respect to the angle $\theta$, which is the angle measured from the negative $y$-axis anticlockwise around the center of the inner circle as shown in Fig.~\ref{fig:noncoaxial_discr}. 
The angle has been taken in this way in order to maintain the geometrical similarity with Ref.~[\onlinecite{Aoki_Sone_1989}].
The unit tangential directions on the inner and outer circles are marked in Fig~\ref{fig:noncoaxial_discr} with blue arrows.
Figure~\ref{fig:vel} shows that the tangential components of the velocity for both inner and outer circles remain zero at $\theta = 0$ and $\theta = \pi$ and that they attain the maximum values somewhere in $(0, \pi/2)$. 
Furthermore, the value of $\theta$ at which the maximum is attained also shifts more toward $\theta = \pi/2$ on increasing the value of the Knudsen number.
Figure~\ref{fig:vel} evinces that the results from the MFS applied on the CCR model (solid lines) are in reasonably good agreement with those from the linearized BGK model for small Knudsen numbers (green lines and symbols) and that the differences between the results from both methods become more and more prominent with increasing Knudsen numbers (red and blue lines and symbols), where the present method starts overpredicting the results, though the trends from both methods remain qualitatively similar to each other even for high Knudsen numbers.
The reason for these quantitative mismatches for large Knudsen numbers is attributed to the limitation of the CCR model in capturing the Knudsen layers, which are more conspicuous near the boundaries for large Knudsen numbers.
The thickness of the Knudsen layers increases with increasing the Knudsen number \cite{Struchtrup2008}, which renders larger deviations in the tangential component of the velocity near the inner and outer walls of the cylinders with increasing the Knudsen number.
%{\color{red}The reason for the differences in tangential velocity results is attributed to the inability of the CCR model to predict the Knudsen layers. 
%The tangential velocity comparison is shown on the inner and outer cylinder boundaries, where the Knudsen layers are prominent than the bulk behavior. 
%As the Knudsen number increases, the thickness of the Knudsen layer increases, resulting in a greater deviation in the velocity profile.}

%The results have been obtained by taking $\Delta\tau=1$ and the parameters $N_b,N_s,r_1,r_2,s_1,s_2,$
%$\alpha_0$ and $\mathrm{Pr}$ have the same values as in the problem discussed in Sec.\,\ref{sec:prob1}.
%In Figure~\ref{fig:vel}, solid lines represent the data obtained via the MFS and the dotted data is the linearized BGK data taken from [\onlinecite{Aoki_Sone_1989}].
%Plots depict the variation of the tangential component of the dimensionless velocity on the inner and outer cylinders. The unit tangent vectors are as indicated in Figure~\ref{fig:noncoaxial} and $\theta$ is the polar angle around the center of the cylinder which is measured from the negative $y$-axis in order to compare the results with [\onlinecite{Aoki_Sone_1989}]. It may be noted that the CCR model overpredicts the tangential velocity  as compared to the linearized BGK model. Nevertheless, it provides qualitatively good results.

\begin{figure}[htb]
\centering
\includegraphics[width=\columnwidth]{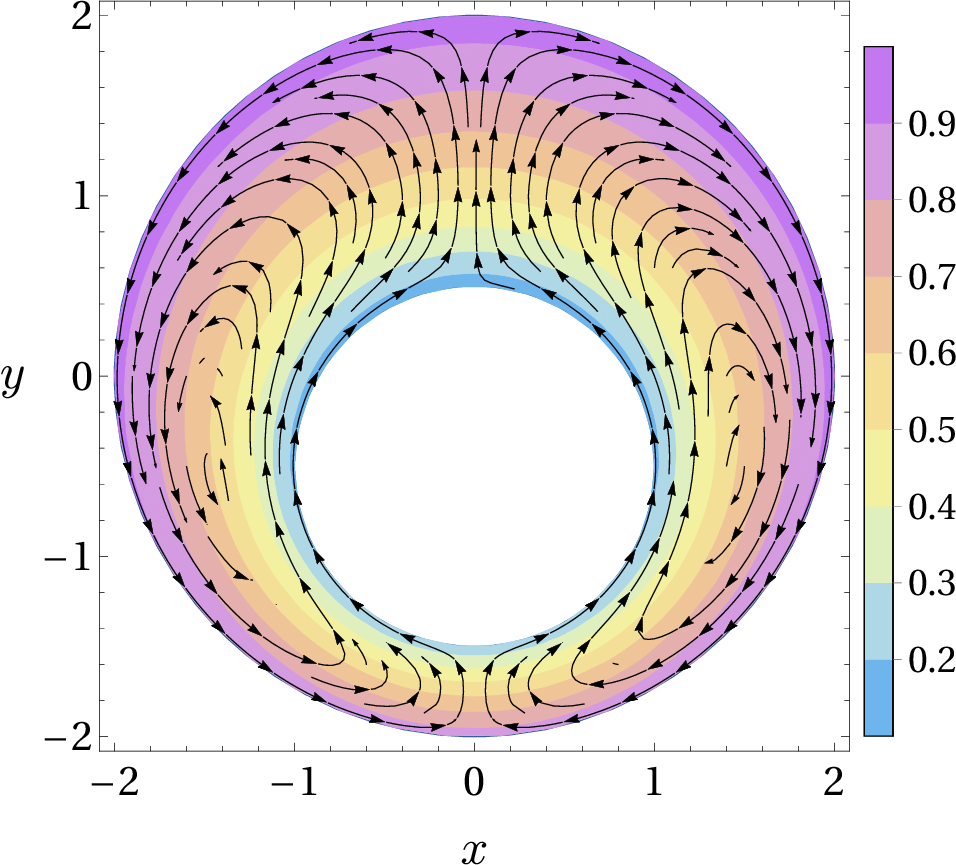}
\caption{\label{fig:stream}% 
Velocity streamlines and temperature contours obtained from the MFS applied on the CCR model for the problem described in Sec.~\ref{sec:prob2} at $\mathrm{Kn}=0.1$ and $\Delta \tau = 1$.
The other parameters are the same as those for Fig.~\ref{fig:temp_gamma}.}
\end{figure}

Figure~\ref{fig:vel}, in other words, also reveals that at $\theta = 0$ and $\theta = \pi$ the flow can happen only in the normal directions.
This prompts us to draw  streamlines of the flow in Fig.~\ref{fig:stream}. 
For explanatory purpose, we also display the temperature contours in Fig.~\ref{fig:stream}.%
\begin{figure}[htb]
\centering
\includegraphics[width=\columnwidth]{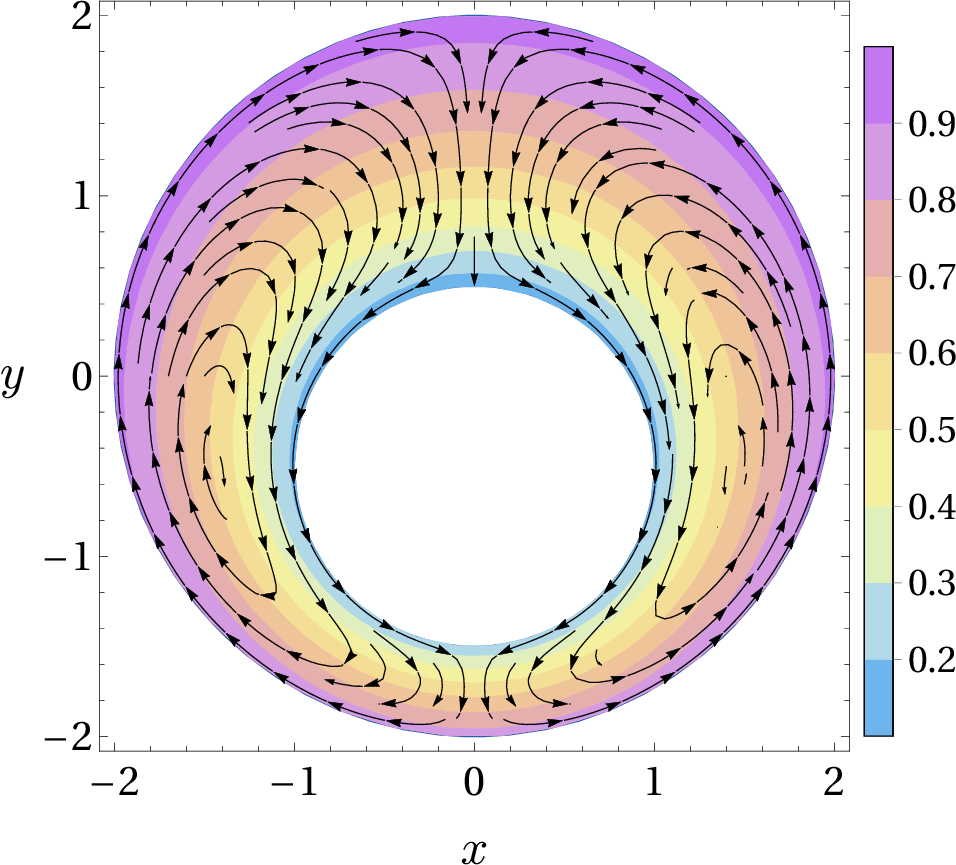}
\caption{\label{fig:NS_stream}% 
Velocity streamlines and temperature contours obtained from the MFS applied on the NSF equations with the second-order slip and jump boundary conditions for the problem described in Sec.~\ref{sec:prob2} at $\mathrm{Kn}=0.1$ and $\Delta \tau = 1$.
The other parameters are the same as those for Fig.~\ref{fig:temp_gamma}.}%
%
%Velocity streamlines predicted by the NSF equations with higher order slip-jump boundary conditions in a cross section for the problem described in Sec.~\ref{sec:prob2} at $\mathrm{Kn}=0.1$.
%The other parameters are the same as those for Fig.~\ref{fig:temp_gamma}.
\end{figure}
The streamlines in Fig.~\ref{fig:stream} show that at the narrowest gap (at $\theta = 0$), the gas starts moving from the outer (hotter) cylinder toward the inner (colder) cylinder due to the largest temperature gradient at $\theta = 0$ and flows along the surface of the inner cylinder on both halves until it reaches $\theta = \pi$, at which it can flow only in the normal direction. 
Therefore, at the widest gap (near $\theta = \pi$), the gas flows from the inner cylinder toward the outer cylinder and returns back from there toward the narrowest gap along the surface of the outer cylinder (but in the opposite directions due to symmetry along the $y$-axis).
This renders two counter-directional circulating flows, one in the left half of the domain and the other in the right half of the domain.
The directions of the circulating flows reverse on taking $\Delta\tau<0$, i.e.~when the inner cylinder is at a higher temperature than the outer one.
With the considered values of the Knudsen number, the directions of the circulating flows apparently do not depend on the Knudsen number.
The direction of the steamlines obtained from the MFS applied on the CCR model in Fig.~\ref{fig:stream} is consistent with that obtained using the linearized BGK model in Ref.~[\onlinecite{Aoki_Sone_1989}].

In order to gain more insight into the process, we have also implemented the MFS to the (linearized) NSF equations [by setting $\alpha_0=0$ in Eqs.~\eqref{ccr_1} and \eqref{ccr_2}] with the second-order slip and jump boundary conditions \cite{Struchtrup2005, ST2008, TTS2009} [obtained by setting $\alpha_0=1/4$ in Eq.~\eqref{bc2} and $\alpha_0^{\prime}=1/5$ in Eq.~\eqref{bc3}], and plotted the streamlines obtained with them in Fig.~\ref{fig:NS_stream}.
From Figs.~\ref{fig:stream} and \ref{fig:NS_stream}, it is evident that, in contrast to the CCR model, the NSF equations even with the second-order slip and jump boundary conditions predict streamlines in completely opposite and incorrect directions. 
This affirms the inadequacy of the NSF equations in describing thermal-stress slip flows \citep{Sone2007} accurately, which---on the other hand---can be described reasonably well with the CCR model due to the coupling between the stress and heat flux.

\begin{figure}[!t]
\centering
\includegraphics[width=\columnwidth]{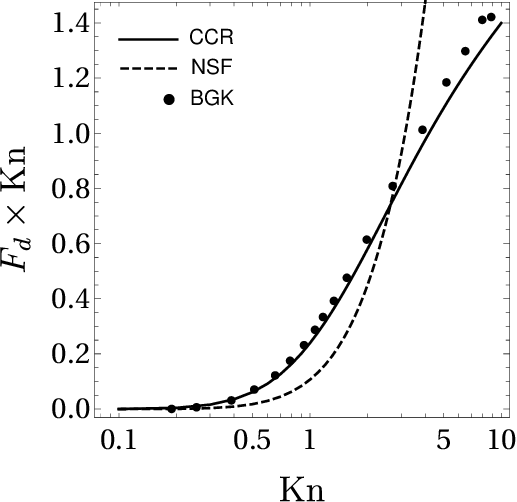}
\caption{\label{drag}%
Drag force on the inner cylinder plotted against the Knudsen number for $\Delta \tau = 1$.
The solid and dashed lines denote the results obtained from the MFS applied on the CCR and NSF models, respectively, while the symbols indicate the data for the drag force obtained from the linearized BGK model \cite{Aoki_Sone_1989}. 
The other parameters are the same as those for Fig.~\ref{fig:temp_gamma}.}
\end{figure}

The superposition of all the point force vectors at the inner source points yields the total force $\bm{F}$ acting on the inner cylinder, i.e.%
\begin{align}
\bm{F}=\sum_{i=1}^{N_s} {\bm{f}}_i,
\end{align}
where $ i=1,2,\dots,N_s$ refer to the points on the inner fictitious boundary. 
The projection of the total force in the direction opposite to the streamwise direction is referred to as the drag force (on the inner cylinder), which is given by 
\begin{align}
F_d = \bm{F} \cdot (-\hat{\bm{y}}) 
= - \sum_{i=1}^{N_s} {\bm{f}}_i\cdot \hat{\bm{y}},
\end{align}
where $\hat{\bm{y}}$ represents the unit vector in the streamwise direction.
% 
%The superposition of the second components of the point force vectors at the inner source points gives the total drag force $D$ on the inner cylinder, i.e.%
%\begin{align}
%D=-\sum_{i=1}^{N_s} {f_2}_i,
%\end{align}
%where $ i=1,2,\dots,N_s$ refer to the points on the inner fictitious boundary and the negative sign represents the direction opposite  to the flow.
%
Variation of the drag force with the Knudsen number is illustrated in Fig.~\ref{drag}, which shows good agreement between the results from the MFS applied on the CCR model (solid lines) and those from the linearized BGK model (symbols) even for high Knudsen numbers (especially, for $\mathrm{Kn}\lesssim 2$).
%The solid lines denote the results obtained from the MFS applied on the CCR model and the symbols indicate the data from the linearized BGK model \cite{Aoki_Sone_1989}. 
This was actually not the case for tangential velocity displayed in Fig.~\ref{fig:vel}, where the differences between the results from the two models were noticeable for high Knudsen numbers.
This shows that the CCR model is capable of predicting the global quantities, e.g., the drag force, quite accurately but is incapacitated of predicting the local quantities, e.g., the velocity and temperature, for high Knudsen numbers due to its limitation of not being able to predict Knudsen layers.
On the contrary, the drag force obtained with the NSF equations (depicted by the dashed line in Fig.~\ref{drag}) deviates significantly from the drag force obtained with the linearized BGK model for $\mathrm{Kn} \gtrsim 0.2$.
%In this plot, The comparison between the CCR model values obtained via the MFS are in good agreement with the linearized BGK model values which is better than comparison in tangential velocity case indicating that the CCR model accurately captures global quantities such as the drag force and heat flow but not the local quantities including velocity and temperature. This occurs due to the inability of the CCR model to predict Knudsen layers.
%
%The drag force acts opposite to flow direction and hence pushes the inner cylinder toward the narrowest gap.

%As illustrated in Figure~\ref{fig:vel}, The MFS-CCR model overpredicts velocity fields but appears to be spot on for drag force as seen in Figure~\ref{drag}. This occurs because the CCR model accurately captures global quantities such as drag force and heat flow but not local values including velocity and temperature. (Knudsen layer??)
\section{\label{sec:location}Location of singularities}

As mentioned in Sec.~\ref{sec:intro}, the collocation matrix associated with the linear system resulting from the MFS could be ill-conditioned and there is a trade-off between the accuracy and good conditioning.
Therefore, it is important to determine an appropriate location for the fictitious boundary in order to obtain the solutions with a desired accuracy.

An ill-conditioned matrix has a high condition number.
Thus the MFS can yield accurate results even with the collocation matrix having a high condition number.
This seems to be implausible intuitively; notwithstanding, it should be noted that the traditional condition number is not adequate for measuring the accuracy and stability of the resulting system since the condition number does not take boundary data into account.
%This is justified as the traditional condition number is not adequate for measuring the stability of the resulting system since the condition number does not take boundary data into account. 
For instance, while forming matrix system \eqref{collocation_system}, the boundary data, such as $p_s$, $\tau_s$ or $\Delta \tau$, for the problems investigated in this paper appear in the vector $\mathcal{B}$ but not in the collocation matrix $\mathcal{M}$. 
Hence, the (usual) condition number of the matrix $\mathcal{M}$ is not an adequate parameter to gauge the sensitivity of the MFS toward the location of the source points.

A more accurate estimation of the sensitivity of the MFS toward the location of the source points can be made by the \emph{effective condition number}, which also takes the boundary data into account (through the right-hand side vector).
The concept of the effective condition number has been used by many authors to determine an optimal location of the singularity points by conjecturing a reciprocal relationship between the inaccuracy of the MFS and the effective condition number \cite{DML2009, WL2011, CNYC2023}. 
%
%Therefore, to get a more accurate estimation of the sensitivity, the effective condition number is used that also takes the right-hand side vector into account.
% The concept of effective condition number was first proposed by Rice 1983 and further improved upon by Chan, Foulser, and Banoczi et al. 
%Various researchers have used the concept of the effective condition number to determine the optimal location of the singularities by conjecturing a reciprocal relationship between the accuracy of the MFS and the effective condition number \citep{DML2009, WL2011, CNYC2023}. 
%
For both the problems discussed in the above sections, we have used the same strategy to place the source points. 
Further details on this strategy are as follows.   

%For matrix system \eqref{collocation_system}, the classical condition number is a measurement of instability based solely on the matrix $\mathcal{M}$, whereas the effective condition number includes both $\mathcal{M}$ and $\bm{\mathcal{B}}$ for more precise measurement.

%It may be noted from the above sections that the elements of the collocation matrix $\mathcal{M}$ is formed using the fundamental solutions based on the distances between the singularity points and boundary nodes whereas the right-hand side vector $\bf{\mathcal{B}}$ is constructed by exploiting the boundary conditions. 

Using the singular value decomposition, $\mathcal{M}$ (having dimensions $n\times m$) can be decomposed as
$\mathcal{M}=U D V^{\mathsf{T}}$, where $U$ and $V$ are $n\times n$ and $m \times m$ orthogonal matrices and $D$ is a $n\times m$ diagonal matrix containing the positive singular values in descending order: $\sigma_1 \geq \sigma_2 \geq \sigma_3 \geq \dots \geq \sigma_r>0$, where $r \leq m$. 
The definitions of the (traditional) condition number $\kappa$ and the effective condition number $\kappa_{\mathrm{eff}}$ in $\ell^2$-norm are given by
\begin{align*}
\kappa=\frac{\sigma_1}{\sigma_r}\quad\text{and} \quad \kappa_{\mathrm{eff}}=\frac{\Vert\bm{\mathcal{B}} \Vert}{\sigma_r \Vert\bm{\mathcal{X}}\Vert}.
\end{align*}
%respectively. 
Using the definition of the effective condition number,  %{\color{red}in which $\Vert\bm{\mathcal{X}}\Vert$ is calculated using the least-square approximation}, 
we first verify the inverse relationship between the maximum error and the effective condition number. 
\begin{figure}[!b]
\centering
\includegraphics[width=\columnwidth]{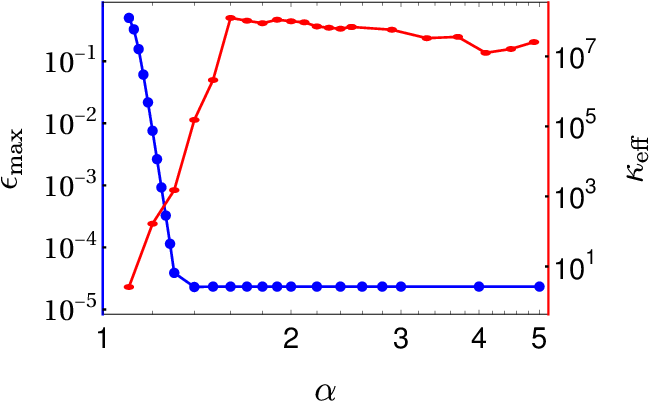} 
\caption{\label{error_keff}
The maximum absolute error $\epsilon_{\max}$ in the temperature and effective condition number $\kappa_{\mathrm{eff}}$ for the problem of flow between coaxial cylinders plotted over the dilation parameter $\alpha$ for $\mathrm{Kn}\approx 0$ and $N_b=N_s=100$.}
\end{figure}
Let $\alpha>1$ be the dilation parameter that determines the separation between the actual boundary (containing boundary nodes) and the fictitious boundary (containing singularities) such that $s_1=r_1/\alpha$ and $s_2=\alpha \,r_2$. 
A larger value of $\alpha$ corresponds to a larger gap between the boundary nodes and source points. 
%A larger value of $\alpha$ corresponds to a larger gap between the actual and fictitious boundaries. 

For the problem described in Sec.~\ref{sec:prob1}, 
the maximum absolute error $\epsilon_{\max}$ in the temperature computed with the MFS and with the analytic solution for $\mathrm{Kn}\approx 0$ along with the effective condition number is plotted against the dilation parameter $\alpha$ in Fig.~\ref{error_keff}. 
The figure shows that the inaccuracy of the MFS is roughly inversely proportional to the effective condition number.
%
%The maximum absolute error the MFS results from the analytical solution for $\mathrm{Kn}\approx 0$ stated in Sec.\,\ref{prob1_disc} has been calculated.  
%Figure~\ref{error_keff} depicts the maximum absolute error and the effective condition number plotted with respect to $\alpha$ showing that the inaccuracy of the MFS is roughly inversely proportional to the effective condition number. 
It is also evident from the figure that the maximum value of the effective condition number is attained for $\alpha$ around $1.6$, where the effective condition number is of order $10^{8}$ and the absolute error is minimum. 
It is worthwhile noting that the order of the effective condition number remains $10^{8}$ for higher values of $\alpha$ beyond $\alpha \approx 1.6$; similarly, the order of the maximum absolute error remains $10^{-5}$ for higher values of $\alpha$ beyond $\alpha \approx 1.6$.

%Therefore, it is preferred to keep the fictitious boundary on the location for which $\alpha\geq 2$, as the error remains constant afterwards and there is a slight decrease in $\kappa_{\mathrm{eff}}$ as well.

\begin{figure}[!tb]
\centering
\includegraphics[width=\columnwidth]{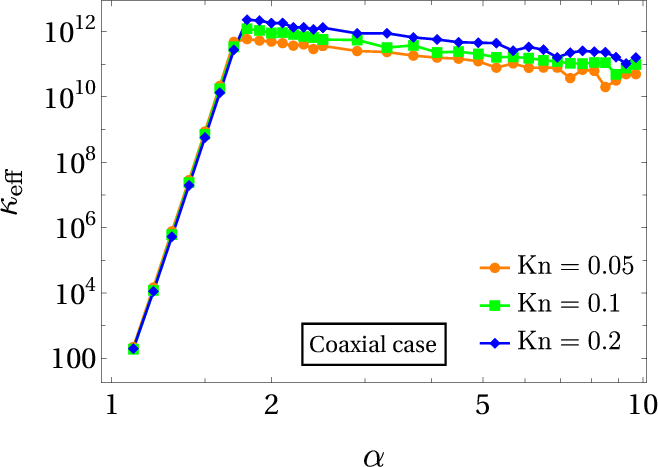}
\\[3ex]
\includegraphics[width=\columnwidth]{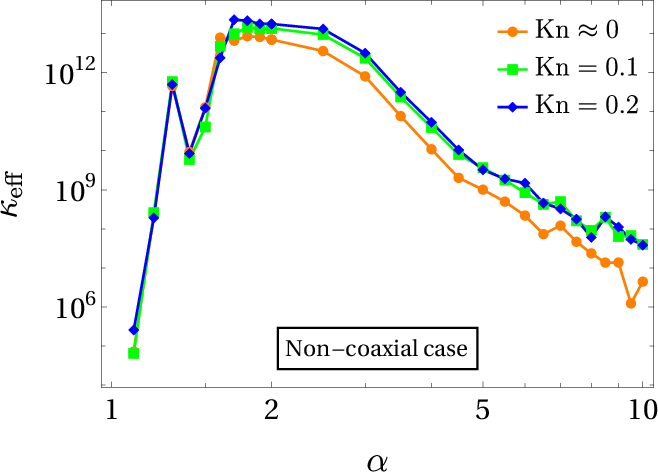}
\caption{\label{keff}
Variation of the effective condition number $\kappa_{\mathrm{eff}}$ with respect to the dilation parameter $\alpha$ for the problems described in Sec.~\ref{sec:prob1} (coaxial case) and Sec.~\ref{sec:prob2} (non-coaxial case). 
The number of boundary nodes at either of the actual boundaries and the number of singularity points at either of the fictitious boundaries  are 100 (i.e.~$N_b=N_s=100$).}
\end{figure}

%In order to have further insights on the effective condition number for the problems considered in Sec.~\ref{sec:prob1} and Sec.~\ref{sec:prob2}, the effective condition number for both problems is plotted against the dilation parameter in Fig.~\ref{keff} for different values of the Knudsen number.
In order to have further insight, the effective condition number for the problems considered in Sec.~\ref{sec:prob1} and Sec.~\ref{sec:prob2} is plotted against the dilation parameter in Fig.~\ref{keff} for different values of the Knudsen number.
The number of boundary nodes at either of the actual boundaries $N_b$ and the number of singularity points at either of the fictitious boundaries $N_s$ are taken as 100 (i.e.~$N_b=N_s=100$) in Fig.~\ref{keff}.
%\begin{figure}[!tb]
%\centering
%\subfloat[coaxial case]{
%\includegraphics[scale=0.65]{figures/eff_cond_coaxial}}\quad
%\subfloat[non-coaxial case]{\includegraphics[scale=0.65]{figures/eff_cond_noncoaxial} }
%\caption{\label{keff}
%Variation of the effective condition number $\kappa_{\mathrm{eff}}$ with respect to the dilation parameter $\alpha$ for the problems described in Sec.~\ref{sec:prob1} (coaxial case) and Sec.~\ref{sec:prob2} (non-coaxial case). 
%The number of boundary nodes at either of the actual boundaries and the number of singularity points at either of the fictitious boundaries  are taken as 100 (i.e.~$N_b=N_s=100$).}
%\end{figure}
%The trends of the effective condition number for different values of $\mathrm{Kn}$ are shown in Figure~\ref{keff} for the problem of coaxial and non-coaxial cylinders discussed in Sec.\,\ref{sec:prob1} and Sec.\,\ref{sec:prob2} respectively. 
It can be noticed from the figure that the highest value of the effective condition number for a given Knudsen number is attained at a value of $\alpha$ somewhere in between $1.8$ and $2$. 
Although, not shown here for succinctness, it turns out that the value of $\alpha$ at which the highest effective condition number is attained increases (decreases) with decrease (increase) in the number of boundary nodes and singularities.
Therefore, to save computational time, one can use smaller number of boundary nodes and source points along with a bigger value of $\alpha$.
From Fig.~\ref{keff}, although the effective condition number decreases on increasing $\alpha$ after a certain value of $\alpha$, we have not encountered any significant change in the results 
%obtained from the MFS 
on keeping the singularities farther  (or on taking bigger values of $\alpha$). 
Therefore, it is apparently sufficient to just ensure $\alpha \geq 2$ to attain an optimal accuracy in the case of $N_b=N_s=100$.
Therefore, the fictitious boundaries for both problems have safely been positioned at locations for which $\alpha = 2$.

%Despite the resultant severe ill-conditioning of the MFS matrix, a minimal impact on the accuracy of solution has been observed. 

\section{\label{sec:conclusion}Conclusion and outlook}
The fundamental solutions of the CCR model in 2D have been determined by exploiting the fundamental solutions of some well-known partial differential equations, e.g.,
the Laplace and biharmonic equations. 
It turns out that the fundamental solutions of the linearized NSF and G13 equations in 2D can be recovered from the fundamental solutions of the CCR model in 2D derived in this paper by taking the coupling coefficient $\alpha_0$ as $0$ and $2/5$, respectively, in them. 
The derived fundamental solutions for the two-dimensional CCR model have then been implemented in a numerical framework.

To gauge the capability of the developed numerical framework, two problems: (i) evaporating/condensing vapor flow between two coaxial cylinders, and (ii) temperature-driven rarefied gas flow between two non-coaxial cylinders having different temperatures, have been revisited. 
These problems have already been investigated with the linearized BGK model in Refs.~[\onlinecite{Onishi1977} and \onlinecite{Aoki_Sone_1989}].
Comparing the results obtained from the MFS for the first problem with those from Refs.~[\onlinecite{Onishi1977}], the accuracy of the MFS with the CCR model in investigating rarefied gas flows with phase change is vivid, particularly for small Knudsen numbers.
%For large Knudsen numbers, the CCR model itself has limitations.
Similarly, for the second problem, the results for the local flow fields, such as the temperature and velocity, obtained using the MFS with the CCR model compares quite well with those obtained using the linearized BGK model in the case of small Knudsen number; but the results for the local flow fields from the two models differ noticeably for larger Knudsen numbers, although their trends from both methods are qualitatively similar.
On the other hand, the MFS with the CCR model is able to capture the global flow fields, such as the drag force, quite accurately even for large Knudsen numbers.
In addition, since the MFS does not involve numerical computation of integrals and its implementation does not require the discretization of the domain, it is computationally efficient in comparison to the other numerical methods used for investigating rarefied gas flows.
This makes the MFS with the CCR model a favorable choice for investigating rarefied gas flows.
It should, however, be noted that  the position of singularity points plays a major role to achieve the best results. 
By performing, effective condition number based studies for both problems, it has been established that the singularity points should be kept sufficiently far from the boundary nodes.
%
%We have determined the fundamental solutions to the CCR model in 2D from which the fundamental solutions of the linearized NSF and Grad-13 models can also be easily seen as a special case by fixing $\alpha_0=0$ and $2/5$, respectively. 
%As applications of the obtained solutions, two cases have been considered that include internal flow problems. 
%In the first case, we have implemented the MFS for phase change problem between two coaxial cylinders and its validation with linearized BGK model results \citep{Onishi1977} shows the efficiency of the two-dimensional MFS in investigating phase-transition flows along with the rarefaction phenomena. In the second case, the MFS for the CCR model has been applied for the temperature-driven flow problem between two non-coaxial cylinders. The comparison of results from [\onlinecite{Aoki_Sone_1989}] shows that the model is qualitatively accurate in producing the thermal stress slip flow but it captures the global flow fields better than the local flow fields. 
%
%In both cases, the position of singularity points plays a major role and to get the best results, the position of singularity points is kept far from the boundary nodes as established from the effective condition number approach.

The utility of the derived fundamental solutions (and their numerical implementation) can be perceived particularly for problems wherein the quasi two-dimensional version of a problem is sufficient to study the complete problem in 3D (due to symmetry in the transverse direction) as the fundamental solutions of a model in 2D cannot be deduced directly from its counterpart in 3D and vice-versa.
Since the derived fundamental solutions in this paper are not restricted to a particular geometry, the findings of this paper also open up the possibility of employing the MFS to quasi two-dimensional problems in complex geometries, for instance, to a flow confined between two right cylinders having non-circular bases.
Furthermore, the derived solutions can also be extended from monatomic to polyatomic gases (of Maxwell molecules) by taking $c_p=(5+\mathsf{n})/2$ and $\alpha_0=2/(5+\mathsf{n})$, and by choosing an appropriate value of the parameter $\mathsf{n}$ that denotes the degree of freedom in a polyatomic gas.

Although the MFS is equally efficient for external flow problems as well (as demonstrated in Refs.~[\onlinecite{LC2016} and \onlinecite{RSCLS2021}]), we have not considered external flow problems in 2D as they are somewhat more involved in comparison to their 3D counterparts. 
This is due to Stokes' paradox in which the solution diverges because of the presence of logarithmic term(s) in the two-dimensional fundamental solutions of Stokes' equation. 
A similar logarithmic term appears in the two-dimensional fundamental solutions of the CCR model as well that makes the study of two-dimensional external flows with the MFS applied on the CCR model involved. 
At present, we do not have a clear understanding of dealing with Stokes' paradox using the MFS for external flow problems, specifically for flows past an arbitrary geometry. 
Notwithstanding, external flows with the MFS applied on the CCR model in 2D will be considered elsewhere in the future.
 % So, in future we shall show the remedy for stokes paradox and apply it for external flows.  
\section{Acknowledgement}
Himanshi gratefully acknowledges the financial support from the Council of Scientific and Industrial Research (CSIR) [File No.: 09/1022(0111)/2020-EMR-I]. A.S.R. acknowledges the financial support from the Science and Engineering Research Board, India through the grants SRG/2021/000790 and MTR/2021/000417. 
Himanshi and V.K.G. also acknowledge the facilities of the Bhaskaracharya Mathematics Laboratory and Brahmagupta Mathematics Library supported by DST-FIST Project SR/FST/MS I/2018/26 that have been used to carry out this work.
\appendix
%\numberwithin{equation}{section}
\section{\label{app:A}Inverse Fourier transforms}
%Derivation of Fourier inverses used to calculate fundamental solutions:
%\[F^{-1}\left(\frac{\bm{\omega\omega}}{\omega^2}\right)=-\frac{1}{2\pi}\left(\frac{2\bm{rr}}{|r|^4}-\frac{I}{|r|^2}\right)\]
%\[F^{-1}\left(\frac{\bm{\omega}}{\omega^2}\right)=-\frac{\mathbbm{i}\bm{r}}{2\pi r^2}\]
%\[F^{-1}\left(\frac{1}{\omega^2}\right)=-\frac{ln|\bm{r}|}{2\pi }\]
%\[F^{-1}\left(\frac{1}{\omega^4}\right)=\frac{r^2(\ln|\bm{r}|-1)}{8\pi }\]
%\[F^{-1}\left(\frac{\bm{\omega}}{\omega^4}\right)=\mathbbm{i}\frac{\bm{r}(ln|r^2|-1)}{8\pi }\]
%\[F^{-1}\left(\frac{\bm{\omega\omega}}{\omega^4}\right)=-\frac{(ln|r^2|-1)}{8\pi}I-\frac{\bm{rr}}{4\pi r^2}\]
%\subsection{\label{sec:f_inv}Inverse Fourier transforms}
We use the fundamental solutions of some well-known equations, such as the Laplace and biharmonic equations, from the literature \citep{Yiorgos2009,PKG1998,CNYC2023} to find the inverse Fourier transforms of the terms on the right-hand sides of Eqs.~\eqref{theta_ft}, \eqref{p_ft} and \eqref{q_ft}--(\ref{sigma_ft}).
Note that the Einstein summation holds over the repeated indices in this appendix and the indices can take values $1$ and $2$ only.
The fundamental solution of the Laplace equation (with a point source of unit strength)  
\begin{align}
\label{laplace}
\nabla^2 \phi \equiv \frac{\partial^2 \phi}{\partial x_i^2} = \delta(\bm{r})
\end{align}%
in 2D is given by
\begin{align}
\phi=\frac{\ln{r}}{2\pi}
\end{align}
where $r=|x_i|$.

Applying the Fourier transformation [defined by Eq.~\eqref{ft}] to the Laplace equation \eqref{laplace}, we obtain
\begin{align}
(-\mathbbm{i})^2 k^2\hat{\phi}=1 \quad\implies\quad 
\hat{\phi}=-\frac{1}{k^2}.
\end{align}
Hence, the inverse Fourier transform of $1/k^2$ is
\begin{align}
\label{app01}
\mathcal{F}^{-1}\left(\frac{1}{k^2}\right)=\mathcal{F}^{-1}(\hat{\phi})=-\frac{\ln{r}}{2\pi}.
\end{align}
Also, by definition~\eqref{ftinv},
the inverse Fourier transform of $1/k^2$ is given by
\begin{align}
\label{app11}
\mathcal{F}^{-1}\left(\frac{1}{k^2}\right)
=\frac{1}{(2\pi)^2} \int_{\mathbb{R}^2} \frac{1}{k^2} \mathrm{e}^{-\mathbbm{i}\, \bm{k}\cdot \bm{r}} \,\mathrm{d}\bm{k}.
%=-\frac{\ln{r}}{2\pi}.
\end{align}
Therefore, from Eqs.~\eqref{app01} and \eqref{app11}, we have
\begin{align}
\label{app12}
\frac{1}{(2\pi)^2} \int_{\mathbb{R}^2} \frac{1}{k^2} \mathrm{e}^{-\mathbbm{i}\, \bm{k}\cdot \bm{r}} \,\mathrm{d}\bm{k}=-\frac{\ln{r}}{2\pi}.
\end{align}
%Also, by definition
%\begin{align}
%\mathcal{F}^{-1}\left(\frac{1}{k^2}\right)&=\frac{1}{(2\pi)^2}\iint\frac{e^{-\bm{\mathbbm{i} k\cdot x}}}{k^2}\,\mathrm{d}\bm{k}
%\end{align}
Now, taking the partial derivative with respect to $x_i$ on both sides in \eqref{app12}, we obtain
\begin{align}
\label{app13}
- \frac{\mathbbm{i}}{(2\pi)^2} \int_{\mathbb{R}^2} \frac{k_i}{k^2} \mathrm{e}^{-\mathbbm{i}\, \bm{k}\cdot \bm{r}} \,\mathrm{d}\bm{k} = -\frac{1}{2\pi} \frac{x_i}{r^2}
\end{align}
which, in turn, gives
\begin{align}
\mathcal{F}^{-1}\left(\frac{k_i}{k^2}\right)&=-\frac{\mathbbm{i} 
\, x_i}{2\pi r^2}.
\end{align}
Moreover, taking the partial derivative with respect to $x_j$ on both sides in \eqref{app13}, we obtain
\begin{align}
\label{app14}
\frac{-1}{(2\pi)^2} \int_{\mathbb{R}^2} \frac{k_i k_j}{k^2} \mathrm{e}^{-\mathbbm{i}\, \bm{k}\cdot \bm{r}} \,\mathrm{d}\bm{k} = -\frac{1}{2\pi} \bigg(\frac{\delta_{ij}}{r^2} -\frac{2 x_i x_j}{r^4}\bigg),
\end{align}
which, in turn, gives
\begin{align}
\label{app3}
\mathcal{F}^{-1}\left(\frac{k_i k_j}{k^2}\right)
=-\frac{1}{\pi}\frac{x_i x_j}{r^4}+\frac{1}{2\pi}\frac{\delta_{ij}}{r^2}.
\end{align}
%
%Indeed,
%\begin{align}
%\label{app2}
%\frac{\partial}{\partial x_i}\mathcal{F}^{-1}\left(\frac{1}{k^2}\right)&=-\frac{\mathbbm{i}}{(2\pi)^2}\iint \frac{k_i}{k^2}e^{-\bm{\mathbbm{i} k\cdot x}}\,\mathrm{d}\bm{k}.
%\end{align}
%%\begin{align}
%%\implies \mathcal{F}^{-1}\left(\frac{k_i}{k^2}\right)&=\mathbbm{i}\frac{\partial}{\partial x_i}F^{-1}\left(\frac{1}{k^2}\right)
%%\\
%%&=-\mathbbm{i}\frac{\partial}{\partial x_i}\left(\frac{\ln{r}}{2\pi }\right)=-\frac{\mathbbm{i} 
%%x_i}{2\pi r^2}
%%\end{align}
%Taking partial derivative with respect to $x_j$ in above equation and following the similar calculations, we obtain
%\begin{align}
%\label{app3}
%\mathcal{F}^{-1}\left(\frac{k_i k_j}{k^2}\right)&=-\frac{\partial^2}{\partial x_j\partial x_i}F^{-1}\left(\frac{1}{k^2}\right)\nonumber
%\\
%&=
%-\frac{1}{2\pi}\frac{2x_ix_j}{r^4}+\frac{1}{2\pi}\frac{\delta_{ij}}{r^2}.
%\end{align}
%Following the same steps for the biharmonic equation 
The fundamental solution of the biharmonic equation (with a point source of unit strength)
\begin{align}
\frac{\partial^4 \phi}{\partial^2 x_i \, \partial^2 x_j}=\delta(\bm{r})
\end{align} 
in 2D is given by
\begin{align}
\phi=\frac{r^2\,\ln{r}}{8\pi}.
\end{align}
%Also,
%We know the fundamental solution for Biharmonic equation from \citep{Yiorgos2009,polyharmonic1983}
%\begin{align}
%\frac{\partial^4}{\partial^4 x_k}\phi=\delta(\bm{r})
%\end{align} 
%is given by
%\begin{align}
%\phi=\frac{r^2\, (\ln{r}-1)}{8\pi}
%\end{align}
%$$\frac{\partial^4}{\partial^4 x_k}\phi=\delta(\bm{r})$$ is $$\phi=\frac{r^2\, (\ln{r}-1)}{8\pi}$$
Following similar steps as for the Laplace equation above, we obtain
\begin{align}
\mathcal{F}^{-1}\left(\frac{1}{k^4}\right)&=\frac{r^2\, \ln{r}}{8\pi},
\\
\mathcal{F}^{-1}\left(\frac{k_i}{k^4}\right)
%&=\mathbbm{i}\frac{\partial}{\partial x_i}F^{-1}\left(\frac{1}{k^4}\right)\nonumber
%\\
&=\mathbbm{i}\frac{x_i(2\ln{r} + 1)}{8\pi },
\\
\mathcal{F}^{-1}\left(\frac{k_i k_j}{k^4}\right)
%&=-\frac{\partial^2}{\partial x_j\partial x_i}F^{-1}\left(\frac{1}{k^4}\right)\nonumber
%\\
&=-\frac{(2\ln{r}+1)}{8\pi} \delta_{ij} -\frac{x_i x_j}{4\pi r^2}.
\end{align}
%
%\section{\label{app:coef}Thermal slip coefficient}
%The expression for the tangential velocity $u_y$ as given in \citep{Sharipov2004} is
%\begin{align}
%u_y=\sigma_P\frac{\mu v_m}{P}\frac{d u_y}{dx}+\sigma_T \frac{\mu }{\rho}\frac{d \ln T}{dy},
%\end{align}
%where $\mu$ is viscosity, $\sigma_P$ is the viscous-slip coefficient and $\sigma_T$ is the thermal-slip coefficient.
%We compare the non-dimensional form of this equation with the tangential component of the velocity given in boundary condition (\ref{bc3}).
%
%Using the relation $\mu=\frac{\mathrm{Pr}}{c_p}\kappa$ and for $\mathrm{Pr}=2/3,\, c_p=5/2$,
%we get
%\[\gamma=\sigma_T\times\frac{4}{15}\]
%Noticing the values for $\sigma_T$ as given in \citep{Sharipov2004}, we get the range of $\gamma$ between $0.2$ and $0.31$.

\bibliography{reference}% Produces the bibliography via BibTeX.

\end{document}